\pdfoutput=1

\documentclass[11pt,twoside,a4paper,note,final]{cms-tdr}
\def\svnVersion{1885a0c-D}\def\svnDate{2019/07/16}

\begin{document}

\hyphenation{had-ron-i-za-tion}
\hyphenation{cal-or-i-me-ter}
\hyphenation{de-vices}

\newskip{\cmsinstskip} \cmsinstskip=0pt plus 4pt
\newskip{\cmsauthskip} \cmsauthskip=16pt

\thispagestyle{empty} 

\vspace*{-25mm}

{\large\sffamily
\hspace*{\fill}
\parbox{45mm}%
{%
{\sffamily         {CERN-LHCC-2019-014}}                  
{\sffamily         {LHCC-I-033}}   \\[2mm]
{\sffamily         {2019/09/09}}       
\\[2mm]
}
}

\vspace*{\fill}

\begin{center}
\Large
{LETTER OF INTENT} \\[4mm]
\LARGE
{\sffamily\bfseries XSEN: a $\nu$N Cross Section Measurement using} \\[3mm]
{\sffamily\bfseries High Energy Neutrinos from pp collisions at the LHC }\\[8mm]
\end{center}

\vspace*{\fill}
\small

N.~Beni$^{1}$$^{,}$$^{2}$, 
S.~Buontempo$^{3}$, 
T.~Camporesi$^{2}$,
F.~Cerutti$^{2}$,
G.M.~Dallavalle$^{4}$, 
G.~De~Lellis$^{2}$$^{,}$$^{3}$$^{,}$$^{5}$, 
A.~De~Roeck$^{2}$,
A.~De~R\'ujula$^{6}$,
A.~Di~Crescenzo$^{3}$$^{,}$$^{5}$, 
D.~Fasanella$^{4}$, 
A.~Ioannisyan$^{2}$$^{,}$$^{7}$,
D.~Lazic$^{8}$,
A.~Margotti$^{4}$,
S.~Lo~Meo$^{4}$$^{,}$$^{9}$, 
F.L.~Navarria$^{4}$,
L.~Patrizii$^{4}$,
T.~Rovelli$^{4}$, 
M.~Sabat\'e-Gilarte$^{2}$,
F.~Sanchez~Galan$^{2}$,
P.~Santos~Diaz$^{2}$,
G.~Sirri$^{4}$,
Z.~Szillasi$^{1}$$^{,}$$^{2}$,
C.~Wulz $^{10}$
\\ [3mm]
\footnotesize
$^{1}$ Hungarian Academy of Sciences, Inst. for Nuclear Research (ATOMKI), Debrecen, Hungary\\
$^{2}$ CERN, CH-1211 Geneva 23, Switzerland \\
$^{3}$ INFN sezione di Napoli, Naples, Italy \\
$^{4}$ INFN sezione di Bologna and Dipartimento di Fisica dell' Universit\`a, Bologna, Italy \\
$^{5}$ Universit\`a di Napoli Federico II, Naples, Italy \\
$^{6}$ Inst. de Estructura de la Materia, Consejo Superior de Investigaciones Cientificas (CSIC), Madrid\\
$^{7}$ A.Alikhanyan National Science Laboratory, Yerevan Physics Institute, Yerevan, Armenia\\
$^{8}$ Boston University, Department of Physics, Boston, MA 02215, USA \\
$^{9}$ ENEA Research Centre E. Clementel, Bologna, Italy \\
$^{10}$ Institut für Hochenergiephysik der OeAW and Vienna University of Technology, Wien, Austria\\

\normalsize

\vspace*{\fill}

\begin{center}
{\sffamily\bfseries Abstract }  \\[2mm]

\parbox[c]{1.0\textwidth}{

XSEN (Cross Section of Energetic Neutrinos) is a small experiment designed to study, for the first time,  neutrino-nucleon interactions (including the tau flavour) in the 0.5-1 TeV neutrino energy range. The detector will be installed in the decommissioned TI18 tunnel and uses nuclear emulsions. Its simplicity allows construction and installation before the LHC Run 3, 2021-2023;   with 150/fb in Run3,  the experiment can record up to two thousand  neutrino interactions,  and up to a hundred tau neutrino events. 

The XSEN detector intercepts the intense neutrino flux, generated by the LHC beams colliding in IP1, at large pseudo-rapidities, where neutrino energies can exceed the TeV.  Since the neutrino-N interaction cross section grows almost linearly with energy, the detector can be light and still collect a considerable sample of neutrino interactions. In our proposal, the detector weighs less than 3 tons. It is lying slightly above the ideal prolongation of the LHC beam from the straight section; this configuration, off the beam axis, although very close to it, enhances the contribution of neutrinos from c and b decays, and consequently of tau neutrinos. The detector fits  in the TI18 tunnel without modifications.

We plan for a demonstrator experiment in 2021 with a small detector of about 0.5 tons; with 25/fb, nearly a hundred interactions of neutrinos of about 1 TeV can be recorded. The aim of this pilot run is a  good in-situ characterisation of the machine-generated backgrounds, an experimental verification of the systematic uncertainties and efficiencies, and a tuning of the emulsion analysis infrastructure and efficiency.

This Letter provides an overview of the experiment motivations, location, design constraints, technology choice, and operation. 
}
\end{center}

\vspace*{\fill}

\parbox[t]{1.0\textwidth}{
\footnotesize
Contacts:~Marco.Dallavalle@cern.ch,~Giovanni.de.Lellis@cern.ch,~Dragoslav.Lazic@cern.ch
}

\parbox[t]{1.0\textwidth}{
\footnotesize
The SHiP Collaboration has provided crucial material help for the
initial measurements and expressed a keen \\
interest in the project
}
\normalsize

\clearpage

\pagestyle{plain}
\setcounter{page}{1}\pagenumbering{roman}

\tableofcontents

\clearpage

\setcounter{page}{1}\pagenumbering{arabic}

\section{Purpose of the Experiment}

\begin{figure}[b]
  \centering
    \includegraphics[width=0.45\textwidth]{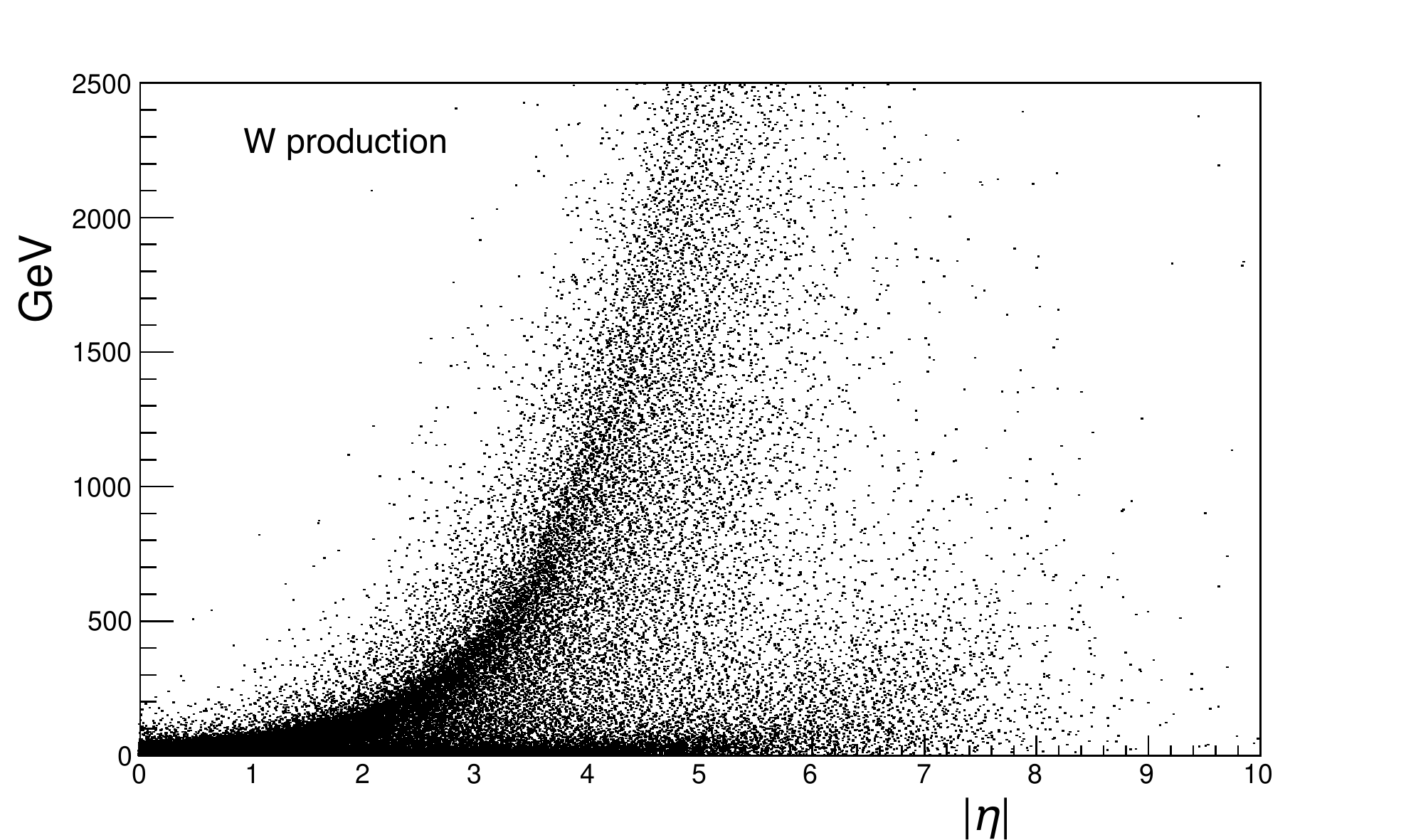}
    \includegraphics[width=0.45\textwidth] {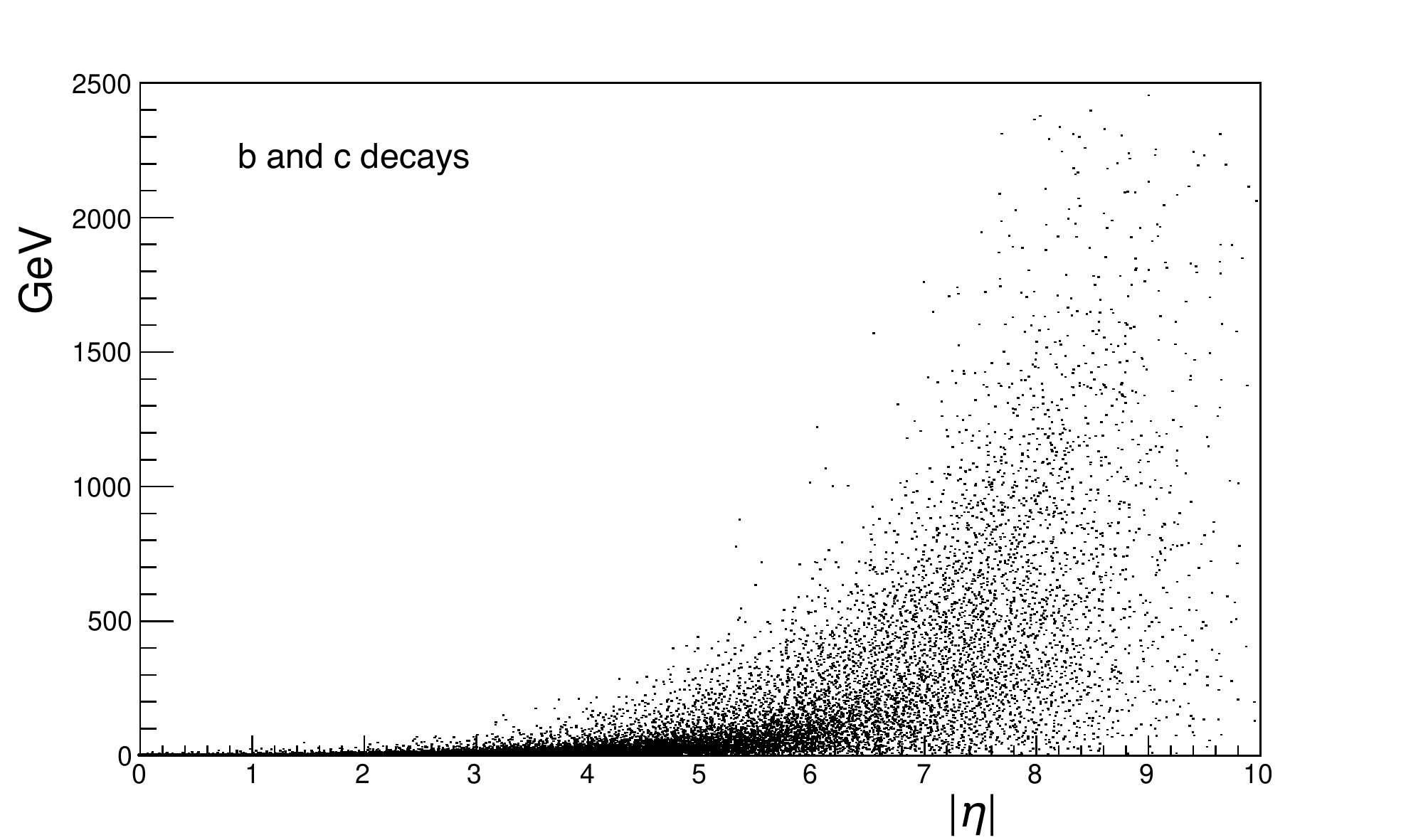}
\caption[Caption scatter plot Wbc] {Scatter plots of neutrino energy versus pseudorapidity $\eta$ 
\cite{XSEN1}. 
Events generated 
with Pythia 8.226 \cite{Pythia}.
Left: pp events with W production; neutrinos from both leptonic and hadronic W decays are shown. Right: neutrinos from b and c decays produced in pp collisions. \label{fig:scatterWbc}}
\end{figure}

The LHC proton beams colliding at  Interaction Points IP1 (ATLAS) and IP5 (CMS) generate intense fluxes of neutrinos of all flavours, including tau
\cite{XSEN1};
at large pseudo-rapidities $\eta$, neutrinos attain TeV energies
(Figure
~\ref{fig:scatterWbc}),
which are a new domain much beyond available neutrino data in accelerator experiments
\cite{PDG}.
At very high energy, astrophysical measurements of the neutrino-nucleon cross section exist
\cite{IceCube}, 
but they are limited to neutrinos of the muon flavour. \\
LHC offers a unique opportunity for  probing the $\nu$N cross section for E$_{\nu}$ larger than 500 GeV and up to a few TeV; tau flavour neutrino interactions are especially interesting, since there are hints of deviations from the Standard Model in the third generation,
from the measurement of the ${\it W}$ decay branching ratio to $\tau$ at LEP
\cite{LEP}
and from measurements of the semileptonic decays of ${\it B}$ to ${\it D}$, ${\it D^{*}}$
\cite{HFLAV}.

In the past two years, we investigated  potential and feasibility of a neutrino experiment at the LHC; 
we focused on high energy neutrinos in two $\eta$ ranges: $4<\eta<5$ ( leptonic W decays,
about 33\% tau neutrinos) and $6.5<\eta<9.5$ (c and b decays, about 5\% tau neutrinos, 
mostly from D$_{s}$ decays);
the results, reported in references
\cite{XSEN1,XSEN2}, constitute the basis for this Letter.

\begin{figure}[t]
  \centering
    \includegraphics[width=0.5\textwidth]{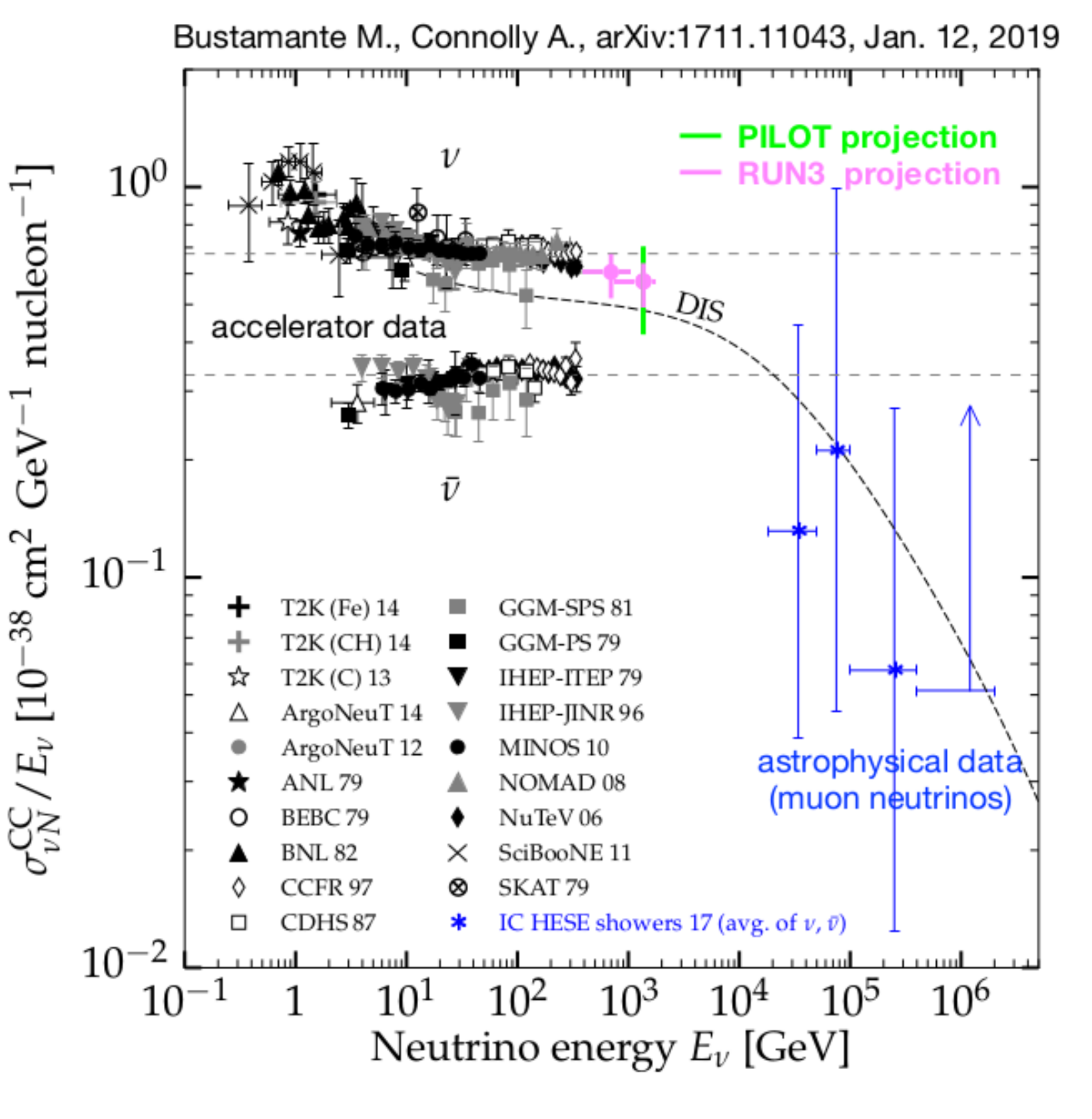}
\caption[Caption Xsection vs E] { XSEN potential measurements (in color), as outlined in 
Table  
~\ref{tab:eventrate},
drawn on a picture from reference 
\cite{xsecvsE} showing existing data. The Pilot point uses only the expected 2021 event statistics.
 \label{fig:xsecvsE}}
\end{figure}

We propose an experiment with acceptance in $7.5<\eta<9.5$, to take data during  the LHC Run3,  with the aim of:
\begin{itemize}
\item
collecting  about hundred  events of tau neutrino interactions. 
\item
measuring the $\nu$N cross section of each flavor 
in two independent energy bins centred at about 0.7 and 1.2 TeV; Figure
~\ref{fig:xsecvsE} shows how the measurements would compare to existing data; 
\end{itemize}

We plan to split the operation into two phases:
\begin{itemize}
\item
Phase 1: from 2019 to 2021. Prepare and perform a pilot run with a detector of about 0.5 ton, aiming at good in-situ characterisation of the machine-generated backgrounds, and a tuning of the emulsion analysis infrastructure and efficiency. 
\item
Phase 2: from 2022 up to the LHC Long Shutdown 3.  A second independent detector is added; the 
target mass extended  to 1.5 and up to 3 tons .
\end{itemize}

We named the experiment XSEN for X-Section of Energetic Neutrinos.

\section{Location and Backgrounds}
 
A first requirement for the experiment feasibility is  a location in which particles coming directly from the IP are screened off, except for  muons and neutrinos,  by rock, or by the absorbers that protect the experimental areas and the LHC components.  
A second requirement is that the local backgrounds from secondary interactions in collimators, beam pipe and other machine elements are low.
Intensity and composition of these machine induced backgrounds vary rapidly along the LHC, however they are well predicted by the simulations performed with Fluka, as extensive in-situ measurements have proven.
\cite{XSEN1, Cerutti1, Cerutti2}.

\begin{figure} [htbp]
\centering
\includegraphics[width=0.7\textwidth]{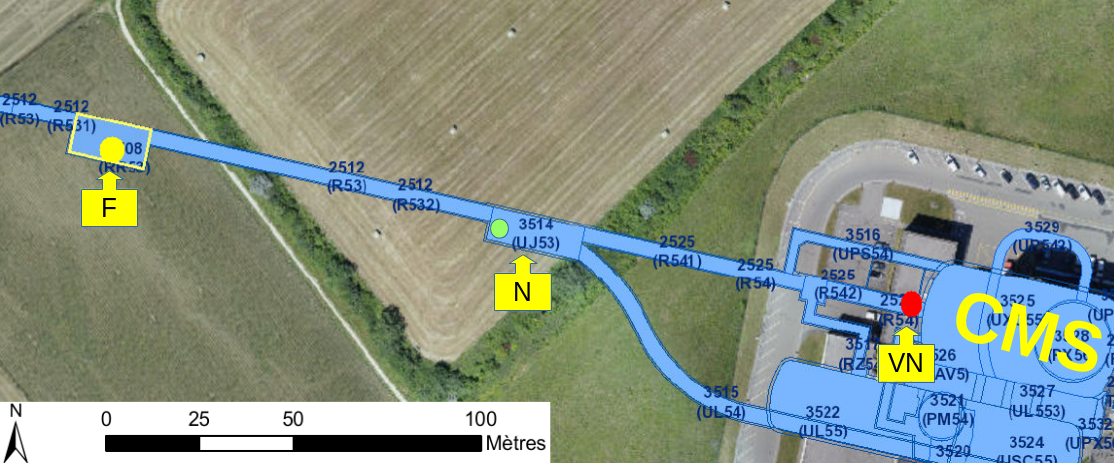}
\caption[Caption emulsions] {View of the VN (Q1 magnet), N (UJ53 hall), and F (RR53 hall) test locations showing their positions and distances from the IP5 along the CMS LHC straight section
\cite{XSEN1}. 
\label{fig:CMS_VN_N_F}}
\end{figure}

\begin{figure} [b]
\centering
\includegraphics[width=0.8\textwidth]{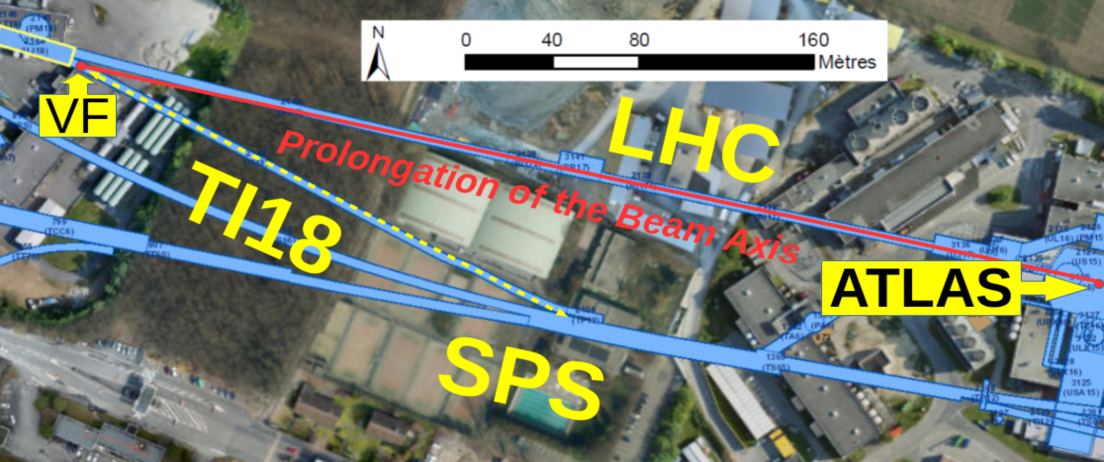}
\caption[Caption emulsions] {View of the VF location showing its position and distance from the IP1. The Prolongation of the Beam Axis (PBA) of the ATLAS LHC straight section intercepts the cavern of the TI18 tunnel after the beginning of the LHC arc
\cite{XSEN1}. PBA is also called Line of Sight (LoS). \label{fig:ATLAS_VF_withPBA}}
\end{figure}

In reference 
\cite{XSEN1}
we compared four locations (named VN, N, F, VF in Figures
~\ref{fig:CMS_VN_N_F} and
~\ref{fig:ATLAS_VF_withPBA})
as potential hosts for a neutrino experiment:
the CMS inner triplet region (~25 m from CMS IP)), UJ53 and UJ57 (90 and 120 m from CMS IP), RR53 and RR57 (240 m from CMS IP), TI18 (480 m from ATLAS IP).
The sites were studied on the basis of
(a) expectations  for neutrino interaction rates, flavour composition and energy spectrum,  (b) predicted backgrounds and in-situ measurements, performed with a nuclear emulsion detector and radiation monitors.

Nuclear emulsions, packaged in stacks with layers of lead as in OPERA
\cite{OPERA}, were used in our tests because they do not require an infrastructure for detector services, and do not disturb the LHC magnets or other machine elements. 

\begin{figure} [t]
\centering
\includegraphics[width=1.0\textwidth]{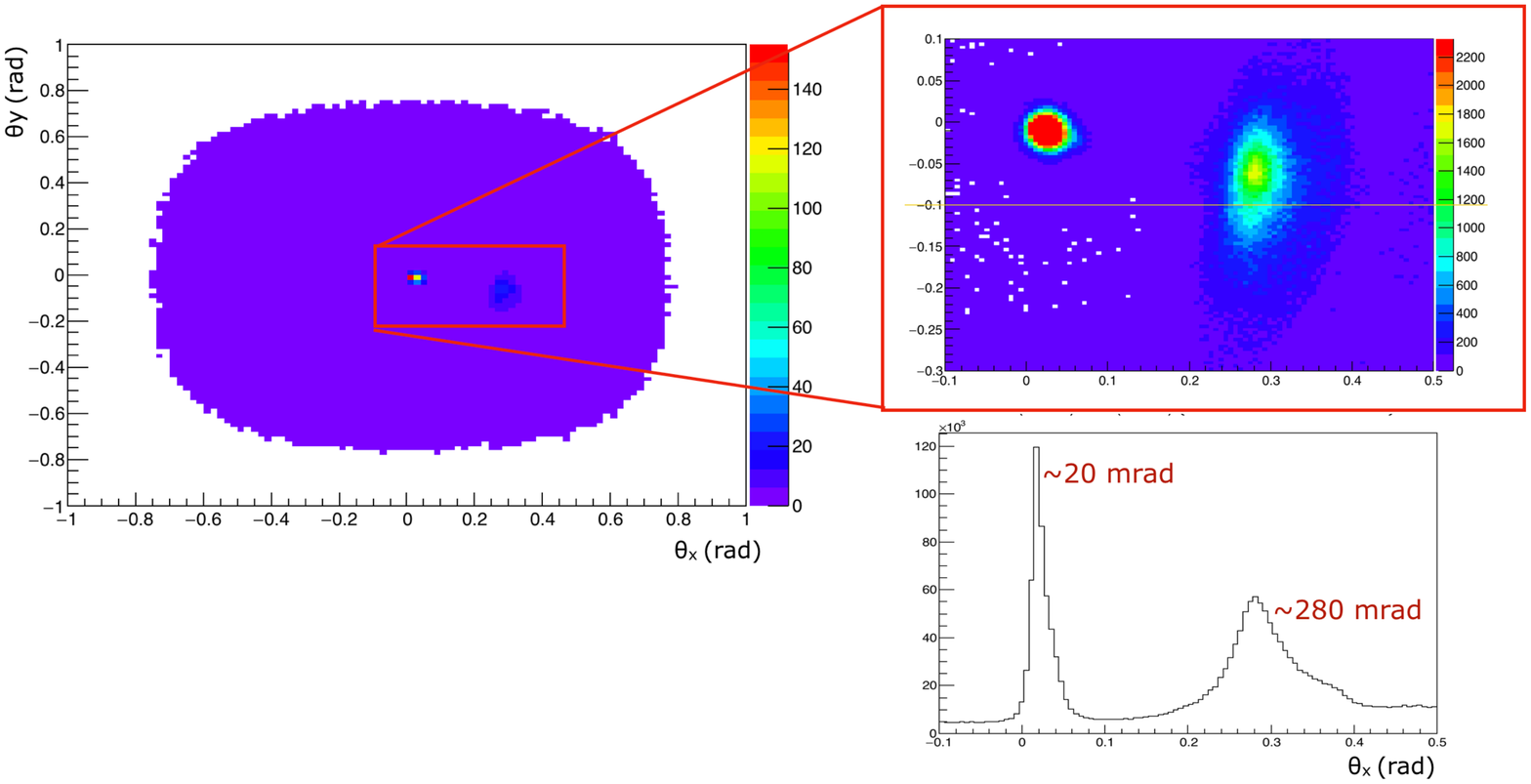}
\caption[Caption scatter plot ADC] {Distributions of the measured inclinations of the observed particle tracks, horizontally ($\theta_{x}$) and vertically, ($\theta_{y}$) in one of the emulsion films of a test package exposed in location F (RR53, 240 m from IP5)
\cite{XSEN1}.
\label{fig:ADC_plot}}
\end{figure}

The emulsions  measured track directions and therefore could separate charged particles coming from the IP
from those  coming from local sources, and from those diffused by tertiary interactions.
Figure
~\ref{fig:ADC_plot} 
 illustrates this with results of the measurements performed in the F location (RR53)
where the emulsion-lead package was exposed and integrated a LHC luminosity of 5.4 fb$^{-1}$.
The sharp peak in the distribution of track angles at ~20 mrad in $\theta_{x}$ and at a few mrad below 0 in $\theta_{y}$ was expected for charged particles coming from the IP. 
The density of the Ag clusters along a track in the peak is compatible with that of muons.
The track population in the peak was measured to be 10$^5$ /cm$^2$, independently in each layer of the package, in good agreement with the expected fluence of muons from the IP in that location.
The peak at about 280 mrad in $\theta_{x}$ and at about -50 mrad in $\theta_{y}$
pointed towards the beam line and indicated a source at a distance 20 meters upstream along the beam line, in both orthogonal measurements:
backgrounds depend on the machine configuration; it was confirmed that during the emulsion exposure, the dominant contribution originated in the nearest quadrupole magnet.
The track population in the 280 mrad peak was determined to be  10$^6$ /cm$^2$. This was consistent with the charged hadron fluence measured with CERN Radiation Monitors
\cite{radmon1, radmon2}, 
that complemented the emulsion package in the background tests.
A background of 10$^7$ /cm$^2$ of tracks at random angles was observed to populate the emulsions uniformly.
The emulsions were protected from thermal neutrons by a 9 cm thick layer of borated polyethylene.  
This was added because it was realized that 
neutrons could excite the silver in a Ag-110m state, a long-lived metastable state produced when a neutron was captured by a Ag-109 nucleus, and electrons and photons from Ag-110m decays would blacken the emulsions. 
In conclusion, tracks of charged particles coming from the IP  could be well identified, although they contributed only about 1\% to the track density in an emulsion. 

 In the three sites tested near IP5 (CMS) (Figure
~\ref{fig:CMS_VN_N_F})
the muon fluence ranged from 1 to 6 $\times$10$^5$  /cm$^{2}$/fb$^{-1}$ in F and VN respectively,
the measured charged hadron fluence was about 10$^6$$-$10$^7$  /cm$^{2}$/fb$^{-1}$, and the thermal neutron fluence was 10$^7$$-$10$^8$ /cm$^{2}$/fb$^{-1}$ in the best location (F)
\cite{XSEN1}.
A luminosity of 1/fb corresponds to roughly 10$^{5}$ seconds at the LHC standard instantaneous luminosity. 
The measurements
were on average in agreement with the predictions 
 of the LHC machine simulations
\cite{Cerutti2}.

\begin{figure}[t]
  \centering
    \includegraphics[width=0.7\textwidth]{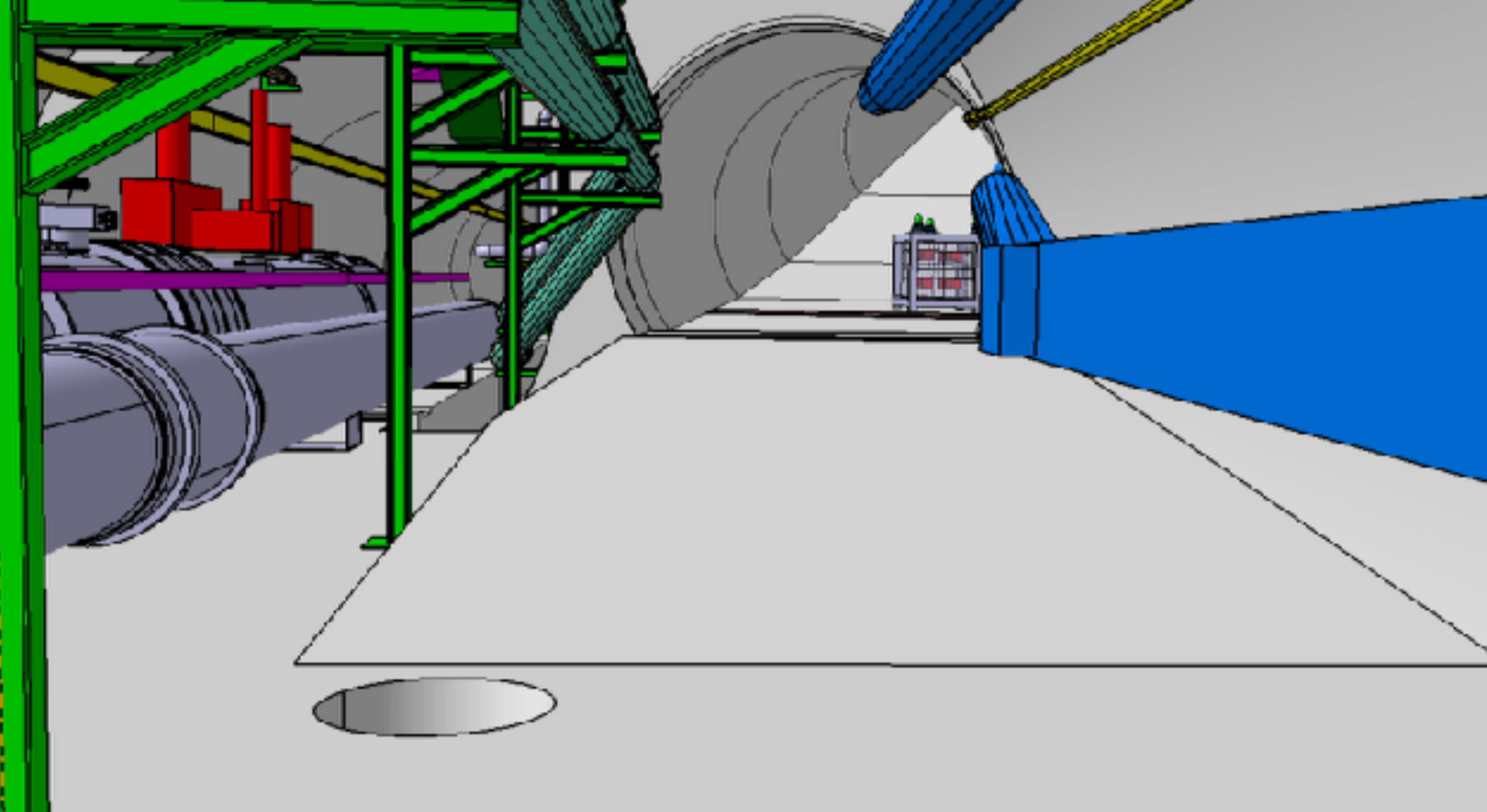}
    \includegraphics[width=0.9\textwidth] {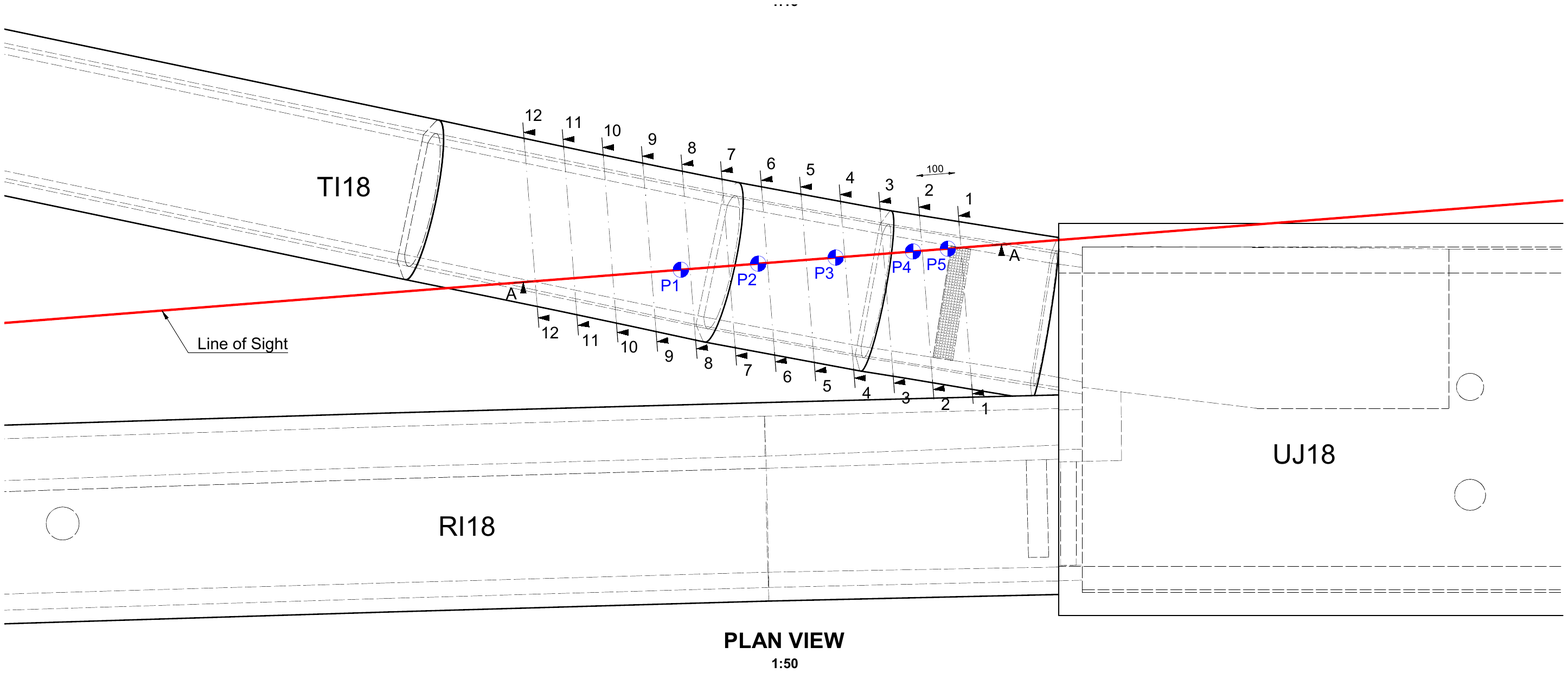}
\caption[Caption TI18] {TI18 tunnel cavern at the connection with the LHC tunnel. Top:  3d view.  Bottom: plan view; IP1 is 480 m on the left. In red the beam Line of Sight from IP1
\cite{CERN-EN}.
 \label{fig:TI18}}
\end{figure}

The VF site only exists near  the IP1 (ATLAS) region
(Figure
~\ref{fig:ATLAS_VF_withPBA}). 
It exploits the decommissioned LEP injection tunnel TI18; a similar tunnel (TI12) exists on the opposite side of ATLAS. 
The TI18 cavern had been considered as location for the FASER experiment 
\cite{FASER}, 
which will be finally installed in TI12, and therefore careful survey measurements were performed by the CERN engineering and survey groups,
and were made available to us
\cite{CERN-EN}.
The cavern of the TI18 tunnel 
(Fig.
~\ref{fig:TI18})
intercepts the prolongation of the beam axis  (named Line of Sight (LoS)), at about 480 meters from the ATLAS IP, at the beginning of the collider's arc, downstream of the first bending dipoles.

The backgrounds in the TI18 cavern (VF location) were not measured by us; they were carefully investigated in preparation for the FASER Technical Proposal
~\cite{FASER}, using
both simulations and in-situ measurements. The analysis of the measurements was refined in a very recent paper
~\cite{FASERnu}.
The equipment was similar to ours: an emulsion-lead package and 
radiation monitors. 
The emulsion exposure integrated a luminosity of 12.5 fb$^{-1}$.
The peak in the track angular distribution pointed to the IP;   2$\times$10$^4$ tracks /cm$^2$/fb$^{-1}$
were counted in the peak, consistent with the expected muon flux from LHC simulation.
The track density in the emulsions was 3$\times$10$^5$ tracks /cm$^2$.
Four BatMon (battery operated RadMon) devices were installed; two were tuned for measuring the high-energy hadron flux, two for measuring the thermal neutron level. The devices were read out after 3 fb$^{-1}$ of LHC luminosity. 
The high-energy hadron fluence was below the device sensitivity of 10$^6$ /cm$^2$. The measured thermal neutron flux was 4$\times$10$^6$ /cm$^2$.

\pagebreak
The measured background levels in VF are an order of magnitude lower than in the F location.
The nuclear emulsions in the F location stood 5.4  fb$^{-1}$.  
In the VF location, the background level would allow for those same emulsions to stand 50-100 fb$^{-1}$ 
and to still remain within the track density limit for analysis. 
We conservately assume 30 fb$^{-1}$ as exposure limit. This is consistent with the luminosity that LHC is expected to deliver in 2021. With these first data we will evaluate the exposure limit in TI18 precisely, and determine the emulsion replacement rate during LHC Run3.   

The muon flux from cosmic rays on our detector in TI18 is very low. The cavern is 80.6 meters underground.  The flux can be evaluated from reference
\cite{cosmics}, considering a standard rock density of 2.65 gr/cm$^3$;
 without restricting the direction of impact on the emulsions, we estimate an upper limit of  10$^3$ tracks/cm$^2$/30fb$^{-1}$.

In summary, the VF location qualified as a suitable host for the neutrino experiment.
However, the area is missing all the basic infrastructures of an experimental zone (lines for power, cooling, and gas systems  etc..).
The useful space is limited in length to a few meters. The floor is on a slope and lies higher than the LoS from a minimum of 5 cm growing in steps up to 50 cm; it can bear a weight of 10 tons/m$^2$.

\section{Detector Design Considerations}

The TI18 location has apparent advantages:
energetic charged particles are deviated by the LHC arc optics and do not reach the detector;
the beam Line-of-Sight from IP1 traverses  about 100 meters of rock before crossing the TI18 cavern.
But, TI18 has important restrictions: it is missing  infrastructures for hosting an electrically active detector with read-out and trigger electronics.
It comes natural to think of using nuclear emulsion packages as done in the OPERA neutrino experiment
\cite{OPERA}.
Emulsions are very efficient for reconstructing the vertex of tau decays, important for tagging tau neutrino interactions.

In our background tests, it was measured that emulsions could stand 10$^7$ tracks
/cm$^2$.
However the track extrapolation between emulsions across a lead layer becomes more complex when the track density is high;
a level of 10$^5$ tracks 
/cm$^2$ is regarded as a good condition.
Thus, given the background estimates,
we prefer not to exceed 30/fb in TI18, although emulsion exposure could last longer. This limitation is to be revisited when
the background levels in the exact detector location are ascertained.
The decision is consistent with the luminosity that LHC is expected to deliver in 2021, while in 2022, 2023 it means replacement of the emulsions at least once during the run, preferably during one of the scheduled Technical Stops.

\begin{figure} [b]
\centering
\includegraphics[width=0.6\textwidth]{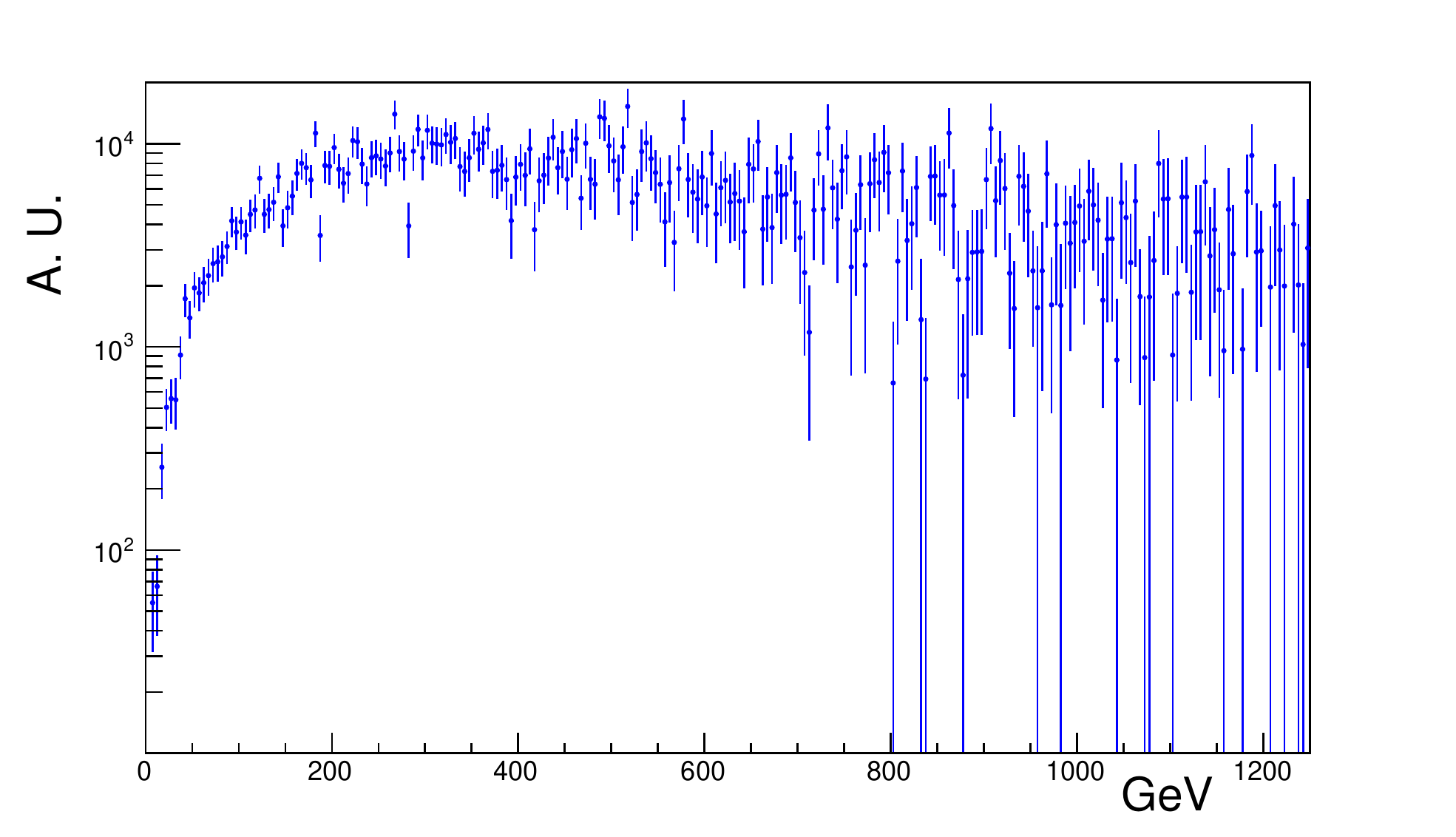}
\caption[spectrum] {Expected energy spectrum for neutrinos interacting in a light mass detector  in the VF location ($\eta>$6.7)
 \cite{XSEN1}.
\label{fig:spectraVF}}
\end{figure}

The $\nu$N cross section grows rapidly with the neutrino energy, linearly from 10 GeV to a few hundred GeV 
(Fig. \ref{fig:xsecvsE}). 
Therefore, at high energy the detector can be light, featuring a mass of a few tons, and still collect a considerable sample of neutrino interactions 
~\cite{XSEN1};
the energy spectrum of the observed events will be hard, because the higher energy neutrinos have larger interaction cross section (Fig.
\ref{fig:spectraVF}).
If lead is used as target, the detector becomes very compact: 
1 ton of lead is a block of 1 meter length and 30x30 cm$^2$ cross section.
The OPERA experiment
\cite{OPERA} used a modular design: emulsions interleaved with thin layers of lead, packed together into  a “brick”.
A brick is 128 mm wide, 104 mm high, and 78 mm thick, with 56 mm lead; it weighs ~8.3 kg.
A 1 ton detector requires 120 such bricks.
This modularity makes easy the installation of the detector, and also its relocation and reconfiguration.
At constant mass, the detector can be made wider and thinner, or narrower and thicker,
which can help in optimizing the event rate
while suppressing lower energy neutrino interactions.

\begin{figure} [t]
\centering
\includegraphics[width=0.7\textwidth]{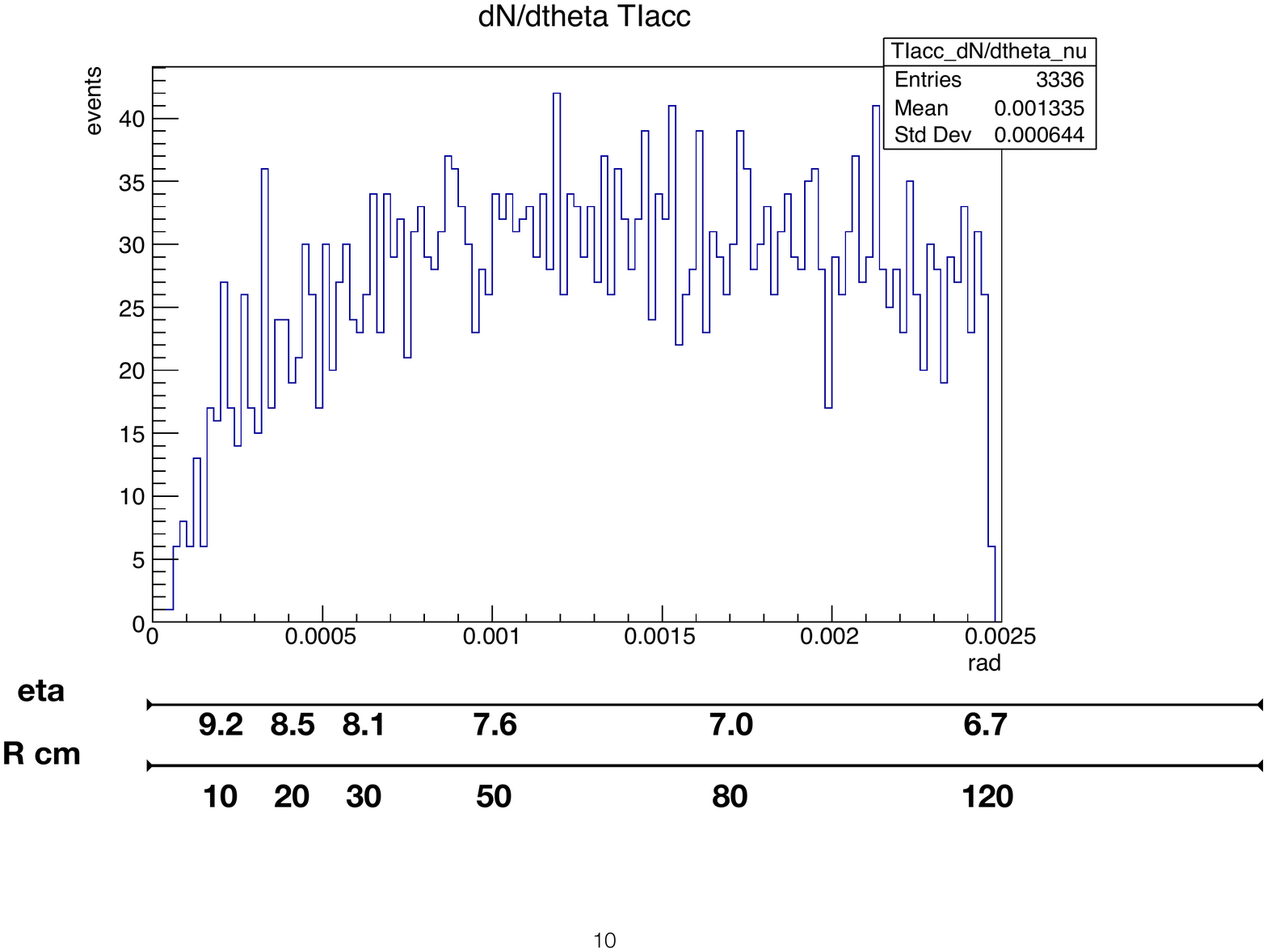}
\caption[neutrinotheta] {Polar angle distribution of neutrinos from b and c decays in the VF location ($\eta>$6.7). Events generated with PYTHIA
\cite{Pythia}.
Also indicated are the pseudorapidity $\eta$ range and the distance R perpendicular to the beam axis.
\label{fig:neutrinotheta}}
\end{figure}

In order to maximize the amount of tau neutrinos, the detector acceptance should favour the contribution from b and c decays.  
Figure 
\ref{fig:neutrinotheta} 
shows the polar angle distribution of neutrinos from b and c decays; about 5\% of those are tau neutrinos.
A configuration with the detector slightly off the beam axis, although very close to it, is preferred.

Emulsion layers in an OPERA brick  are uniformly spaced: a one millimiter lead sheet separates two consecutive layers; there are 60 layers.   We can take advantage of the longer mean path of particles in a high energy interaction and modify the brick structure. 
We are investigating the possibility of increasing (~factor two to three) the longitudinal thickness of the brick in order to improve the tau lepton decay identification efficiency. The expected decay length is indeed much longer than in the OPERA case. 
Dedicated light-tight boxes made of plastic material will be used in this case.
Furthermore, groups of layers (trigger layers) having emulsions close to each other ( interleaved with 1 mm lead sheets ) can be put in four key positions along the brick, separated by regions where the emulsions are interleaved with thicker lead sheets.
The trigger layers are examined at first, for detecting the presence of an event and to start tracking. 
The emulsion analysis stage becomes lighter and quicker.
Of course the brick structure strongly depends also on the backgound track density. Further developments of this idea are planned.
 
\begin{figure} [b]
\centering
\includegraphics[width=0.6\textwidth]{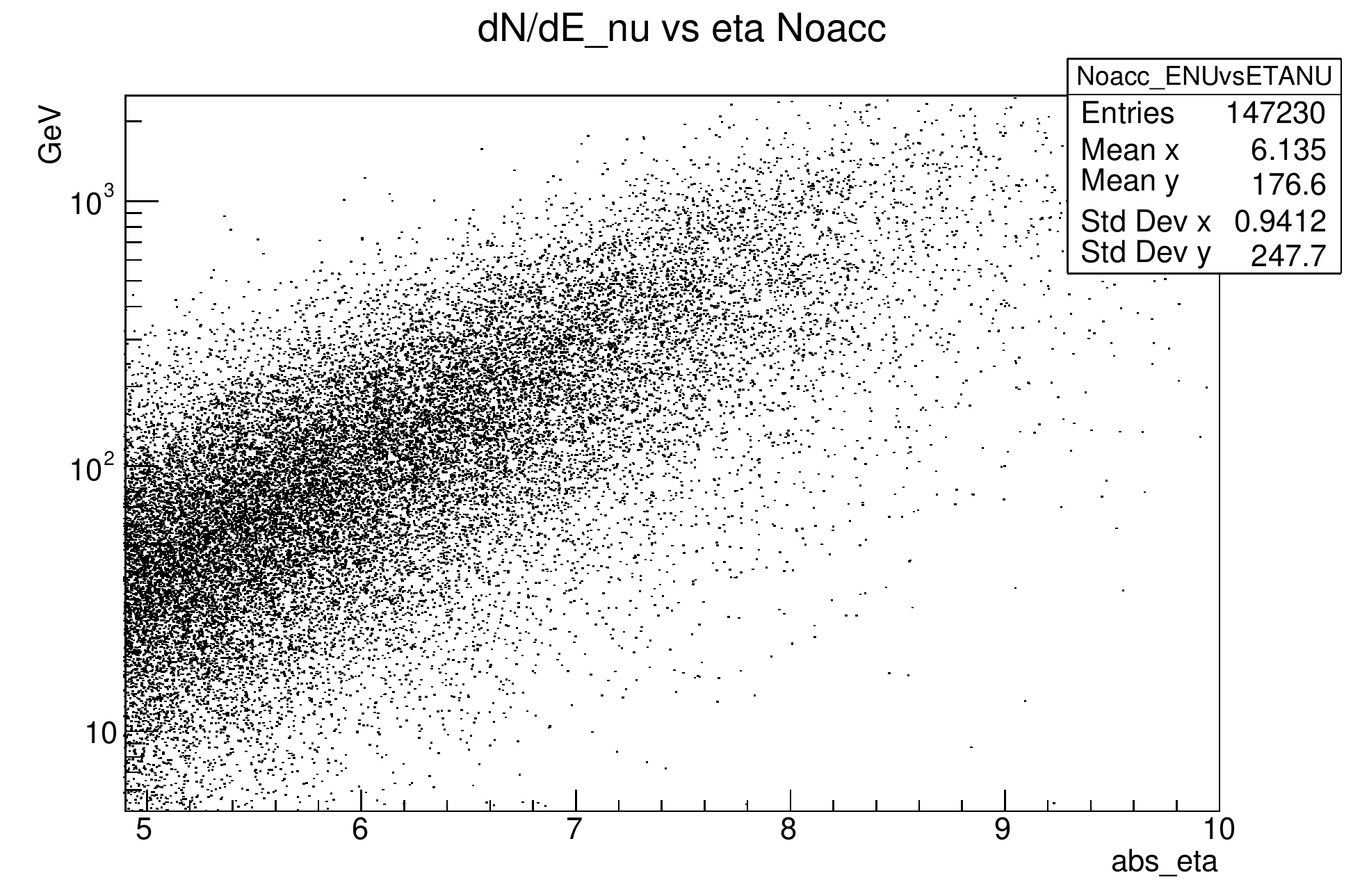}
\caption[neutrinotheta] {A scatter plot of log(E) versus $\eta$ for neutrinos from b and c decays.
Events generated with PYTHIA
\cite{Pythia}. 
\label{fig:logE_vs_largeeta}}.
\end{figure}

The emulsion packages are very good for reconstructing the vertex of a neutrino interaction, but they cannot easily provide a measurement of the neutrino energy. 
On an event by event basis the neutrino energy can be estimated using the methods developed in OPERA and other emulsion based detectors, which for instance exploit the correlation between track multiplicity and neutrino energy
~\cite{OPERA_MCSrec, OPERA_EMrec, SHIP_EMrec}. 
The achievable resolution depends on the brick structure which is begin optimized; we target a resolution of  50\%  in the 0.1-1 TeV energy range.
In addition, kinematics can be exploited: in the regime of longitudinal momentum p$_{L}$ much larger 
than transverse momentum p$_{T}$, the pseudorapidity of particles emerging from the IP is proportional to the logarithm of the energy, 
a relation smeared by the particle p$_{T}$ distribution.
Different η ranges have different average energy, as seen in figure
\ref{fig:logE_vs_largeeta}.
We propose a design with two independent detectors, covering different $\eta$ ranges, and
we do not aim at a high energy resolution, but rather to have 2 bins, one for each detector.
The  log(E) versus $\eta$ scatter plot can also be useful for defining a fiducial phase space so to reject obvious outliers.

The LHC simulations can predict the neutrino flux from pp collisions and its composition, so that the expected event rate in the detector acceptance can be directly compared to the measurement, without separating neutrino and anti-neutrino. 
However, if a deviation is observed we can tell if it is due to a specific flavour: since neutrino flavours are identified in the detector by event chracteristics, flavour ratios can be measured.

In XSEN the emulsion exposure in TI18 can last four months or longer, depending on machine backgrounds, before
replacement.
Hence, an important parameter is the fading time of an event recorded in an emulsion.
Altough there are many emulsion layers in a brick, a short fading time can create analysis inefficiencies. 
Investigations  are under way, qualifying tests are foreseen.

\section{Neutrino Flux}

Pythia 8.226
\cite{Pythia}
was used in reference
\cite{XSEN1}
to simulate proton-proton collision events at $\sqrt{s}=$ 14 TeV and to estimate the neutrino flux.
Most of the high energy neutrino flux in the range $6.7<\eta<9.0$ originate from  c and b decays; about 5\% of the neutrinos are of the tau flavour.  

Pythia was tested in depth by the LHC experiments; it reproduces the features of proton-proton interactions with good accuracy.
Measurements of charged particle production in the forward direction were performed by LHCb
\cite{LHCb_charged} in the pseudorapidity range $2.0<\eta<4.5$, and by TOTEM
\cite{TOTEM_charged}
 in  $5.3<\eta<6.5$, and found reasonable agreement with Pythia expectations. \\
However no crosschecks exists for the very forward $\eta$ range that XSEN will subtend;
therefore in order to evaluate the reliability of estimates using Pythia, a comparison with published calculations by
A. De R\`ujula, E. Fernandez and J.J. G\`omez-Cadenas 
\cite{derujula} (1993)
and by H. Park
\cite{Park}
(2011)
was performed.
De R\`ujula et al. made an analythical calculation using two variants of a QCD non-perturbative model (Quark Gluon String Model).
Park used a Pythia generator version 6 with parameter set tuned to Tevatron data.
For comparing the results, the rate of neutrino charged current interactions was normalised to the same acceptance, target mass and LHC luminosity:
the upper half of a cylinder of radius 1.2 m with axis on the beam LoS at 500 m from the IP, a thickness of 2 meters, made of lead, standing for 100 /fb.
The results are reported in Table
 ~\ref{tab:fluxcomparison}.  
\begin{table} [h]
\begin{center}
\topcaption{ Expected CC interactions in the upper half of a cylinder of radius 1.2 m with axis on the beam LoS, 2 meters long, at 500 m from the IP, made of lead, standing for a LHC luminosity of 100 /fb, extrapolated from references 
\cite{derujula} (De R\`ujula et al. 1993),
\cite{Park} (Park 2011) and
~\cite{XSEN1}
(Beni et al. 2019). }
\label{tab:fluxcomparison}
\begin{tabular} {lrrrr} \hline 
& & & & \\
expected CC interactions  &De R\`ujula et al.  &De R\`ujula et al. &Park &Beni et al. \\
(neutrino + antineutrino) &1993 variant1 &1993 variant2 &2011  &2019 \\ \hline
& & & & \\
all flavours &2650 &3450   &33518  &9240   \\
tau flavour   &237 &361   &67  &287  \\ \hline
& & & & \\
\end{tabular}
\end{center}
\end{table}
The lower expectation of tau neutrino interactions by Park is due to a lower \textit{D$_{s}$} production cross section.
The difference in the total number of neutrinos between Park and Beni et
al. 
\cite{XSEN1}  
comes from low energy muon neutrinos very close to the beam axis. 
Although these neutrinos are mostly outside the XSEN acceptance, further investigations are planned.
The difference between Beni et al. and De Rujula et al. for all flavour is within the large uncertainty of the QGSM model in predicting the transverse momentum distribution for the produced hadrons.

\begin{figure}[htbp]
  \centering
    \includegraphics[width=0.45\textwidth]{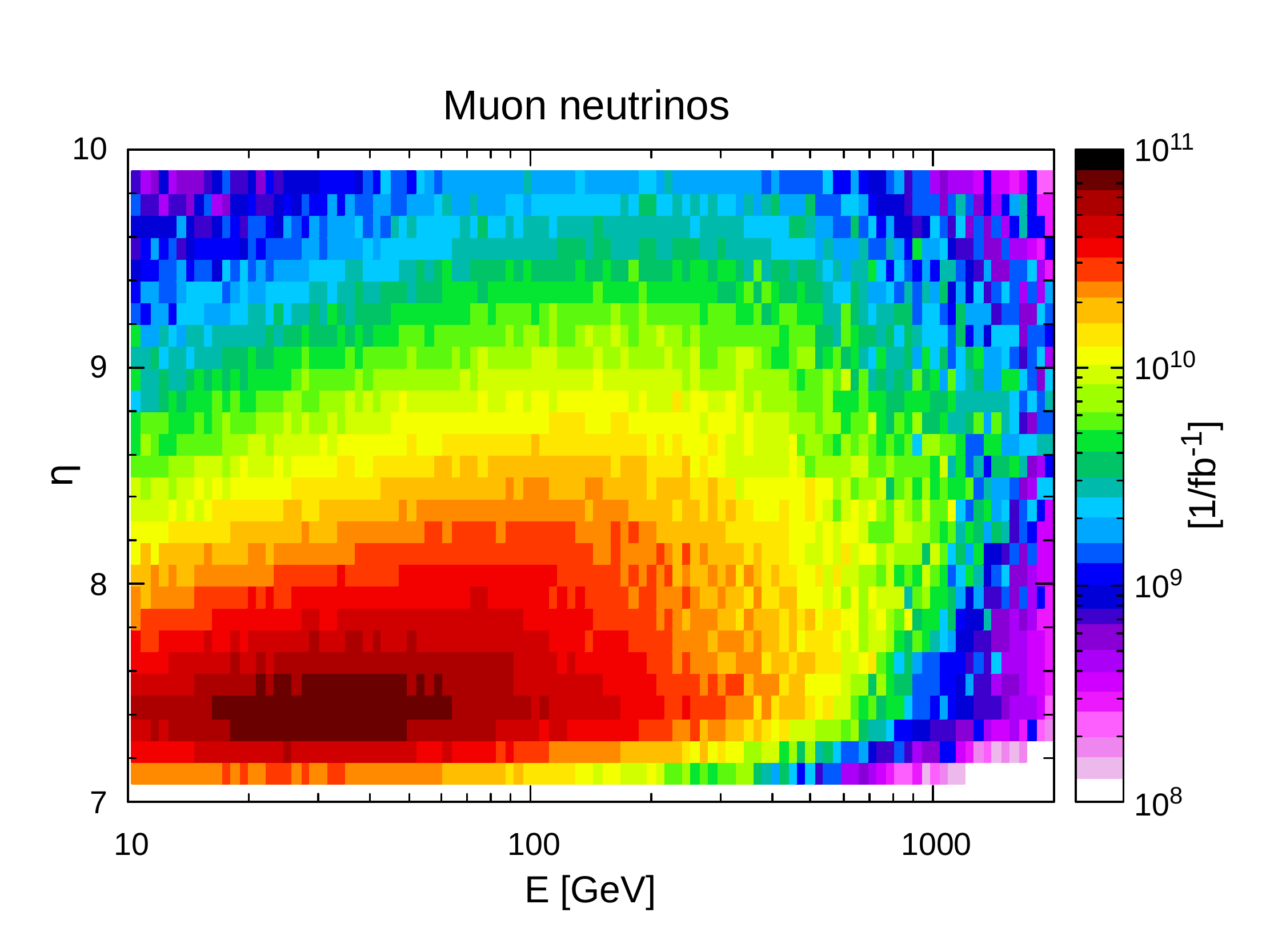}
   \includegraphics[width=0.45\textwidth] {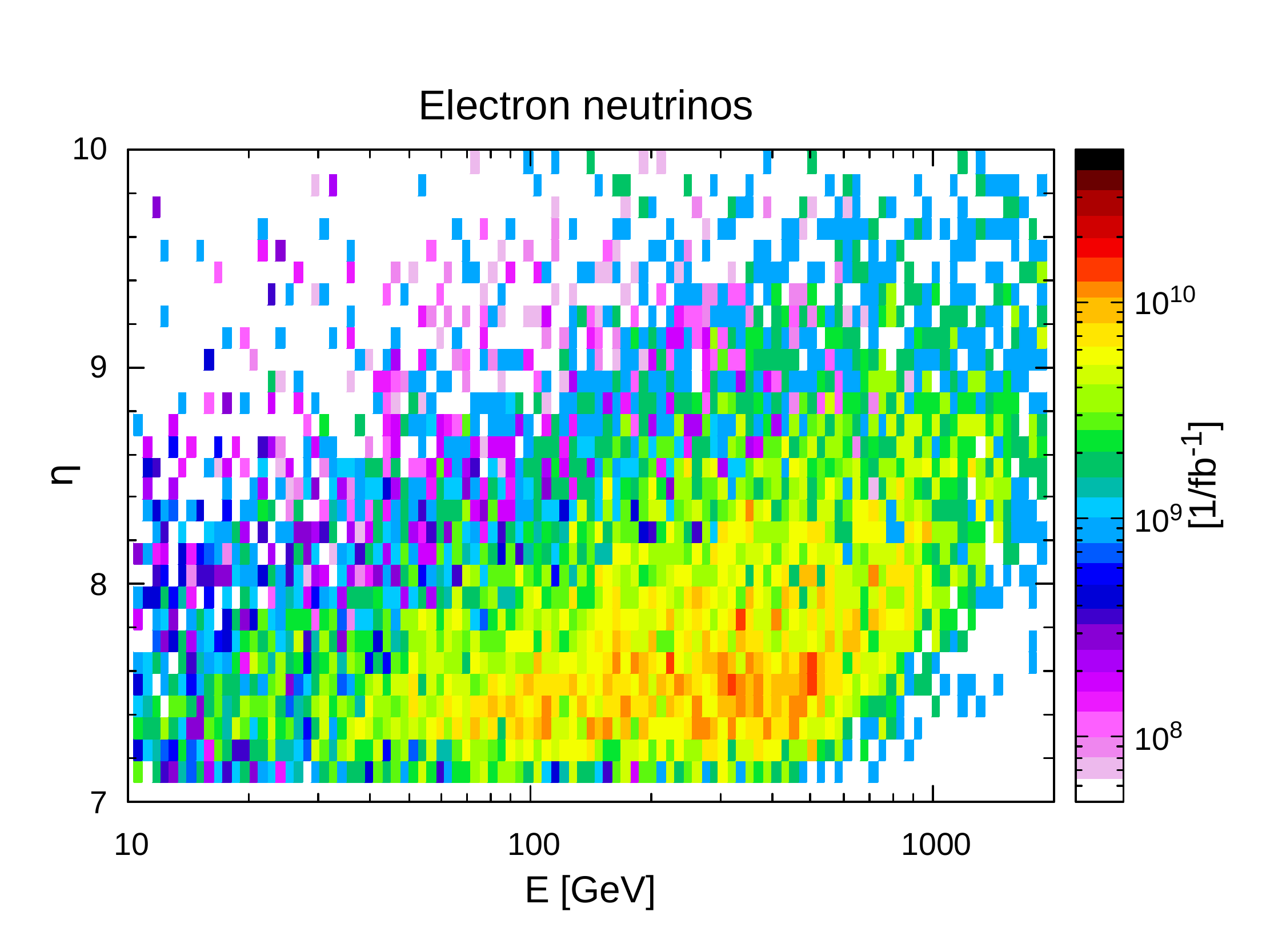}
\caption[Caption DPMJET_scatter] {Scatter plots of neutrino pseudorapidity $\eta$ versus energy. 
Events were generated with Fluka 
using the embedded DPMJET event generator, and LHC simulation; both pion/Kaon decays and charm production were included.
Left: muon neutrinos. Right: electron neutrinos. \label{fig:DPMJET_scatter}}
\end{figure}
\begin{figure}[htbp]
  \centering
    \includegraphics[width=0.5\textwidth]{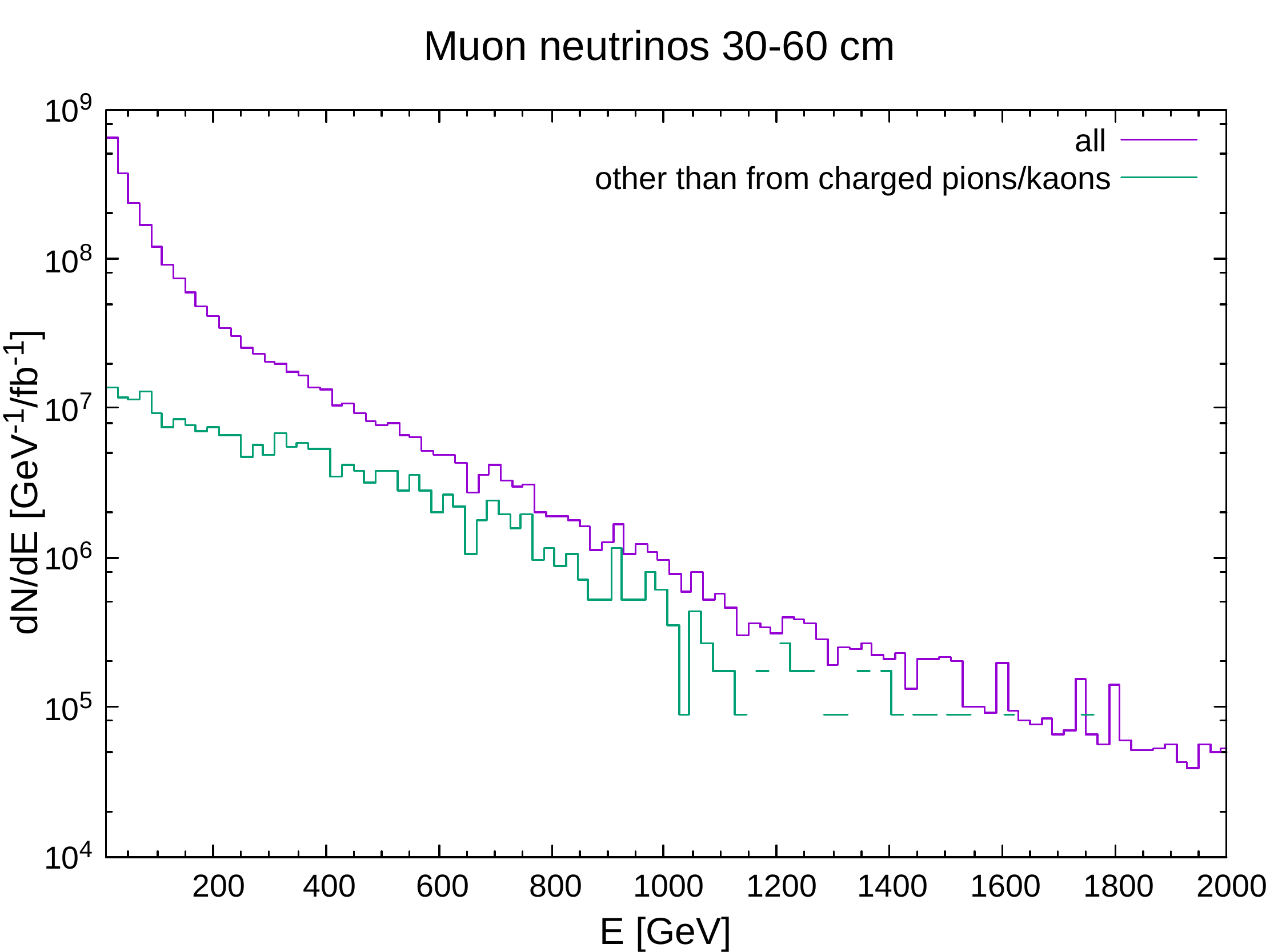}
    \includegraphics[width=0.5\textwidth] {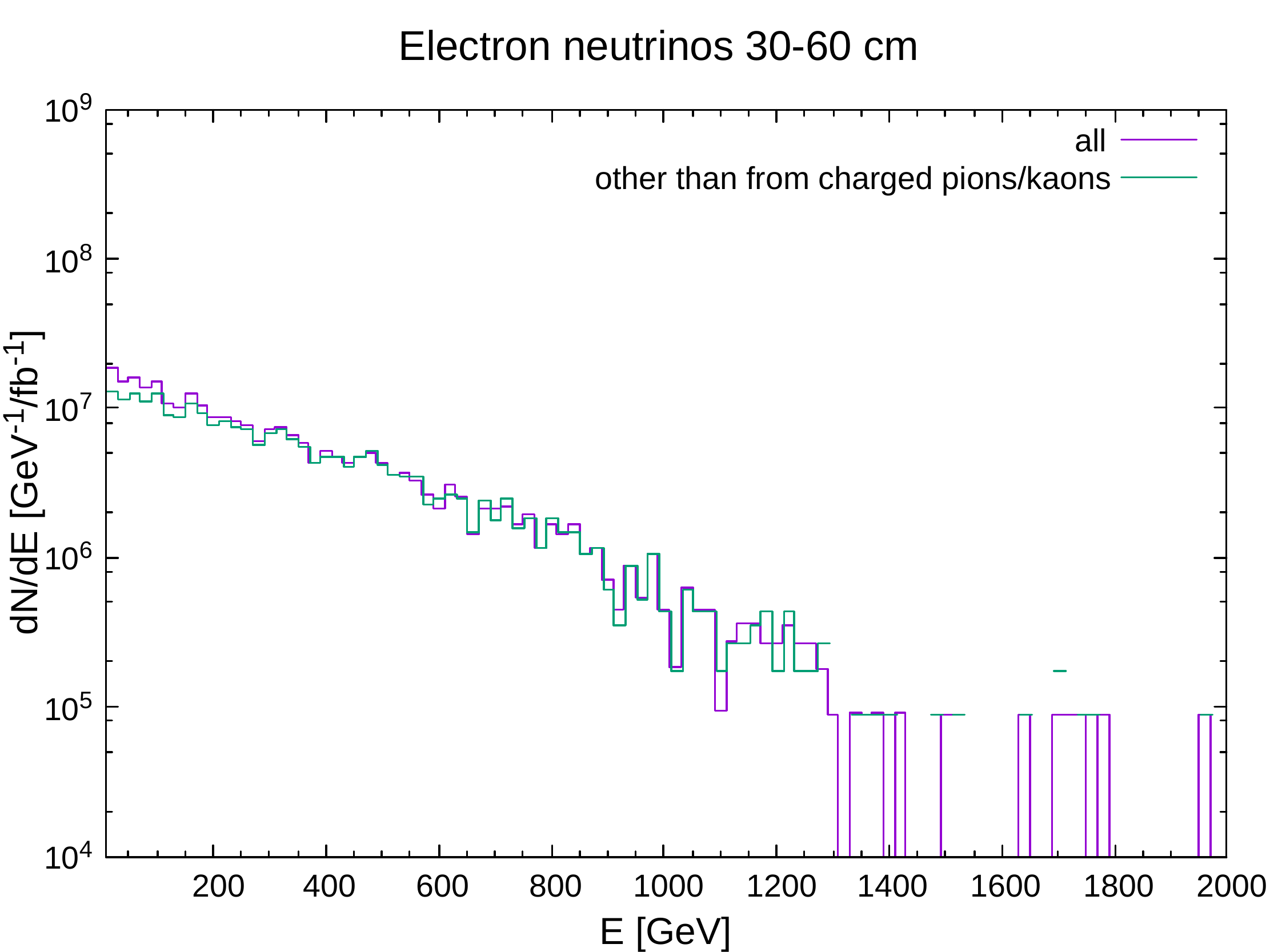}
    \includegraphics[width=0.5\textwidth] {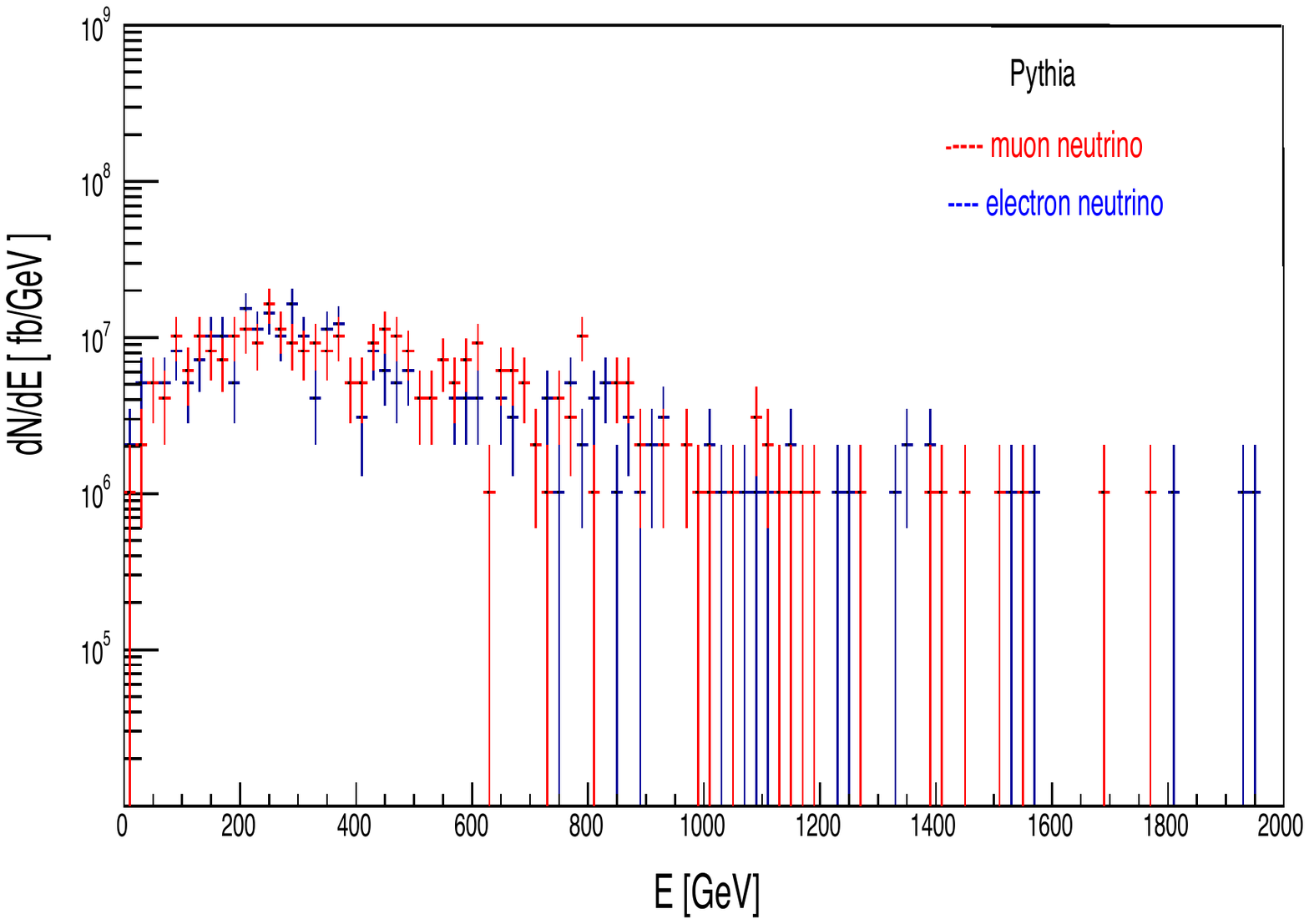}
\caption[B1B2 architecture] {Predicted fluxes of neutrinos for the radial distance $30<R<60$ cm from the beam centre in TI8.
Top: muon neutrinos from DPMJET (Dual Parton Model, including charm) and LHC simulation. 
Middle: elecron neutrinos from DPMJET and LHC simulation. 
Bottom: muon and electron neutrinos from Pythia (charm production). 
 \label{fig:fluxes}}
\end{figure}

Neutrinos from pion decays pointing towards TI18 are predicted to have mostly low energies: pions of 10 GeV have a $\gamma$c$\tau$ of about 550 m, pions of 100 GeV 5.5 Km; therefore most of high energy  pions are deviated by the LHC optics and interact in the LHC beam pipe or in the rock before they can decay.
Neutrinos from Kaon decays pointing towards TI18 can have higher energies, 
the $\gamma$c$\tau$  for a 100 GeV Kaon being about 740 m, 
however the Kaon/pion production rate ratio is only about 11\%.
A thorough study of the expected flux in TI18 very close the beam LoS 
(within 10 cm around it, i.e. $\eta>9.1$ )  was presented in a very recent paper by the FASER collaboration
\cite{FASERnu}. 
Simulation studies are ongoing to better understand how this additional contribution from pion and Kaon decays enters in our specific case, since our detector covers a different eta range and  is less sensitive to low energy neutrinos. 
The CERN FLUKA team in the EN-STI group is very actively pursuing this topic. 
Proton-proton collisions were generated with Fluka 
using the embedded DPMJET event generator, which  describes soft multiparticle production, including charm production;
then pions and Kaons, before decaying, were transported through LHC elements and environment material up to TI18
\cite{Fluka1, Fluka2, DPMJET, LHCsim}.
Information was stored separately for neutrinos and antineutrinos, and by flavor.
Figure ~\ref{fig:DPMJET_scatter} shows the predicted fluence in $\eta$ versus energy, for the eta range under consideration, for muon and electron neutrinos, in 1/fb. 
In figure  ~\ref{fig:fluxes}
neutrino energy distributions are plotted for the radial distance $30<R<60$ cm from the beam centre in TI8, corresponding to $7.4<\eta<8.1$, separately according to whether neutrinos originated from pion/Kaon decays or not. The plots are consistent with the expectations that electron neutrinos and high energy muon neutrinos do not come from pion/Kaon decays.
In figure  ~\ref{fig:fluxes} also the flux predicted with Pythia in the same $\eta$ range of muon and electron neutrinos from charm production in LHC pp collisions is shown; the comparison supports our interpretation. Studies are going on.

\begin{figure}[htbp]
  \centering
    \includegraphics[width=0.6\textwidth]{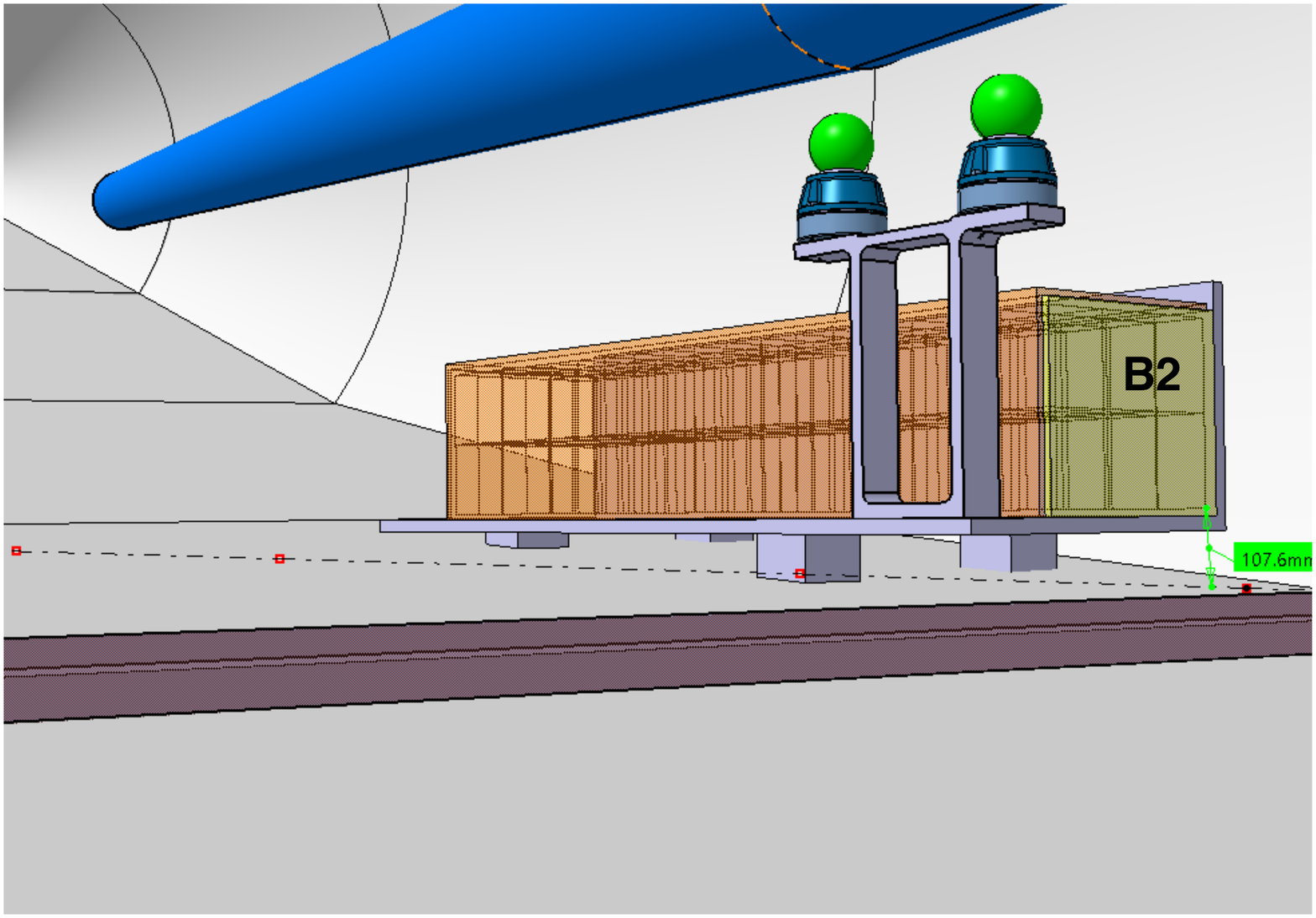}
    \includegraphics[width=0.45\textwidth] {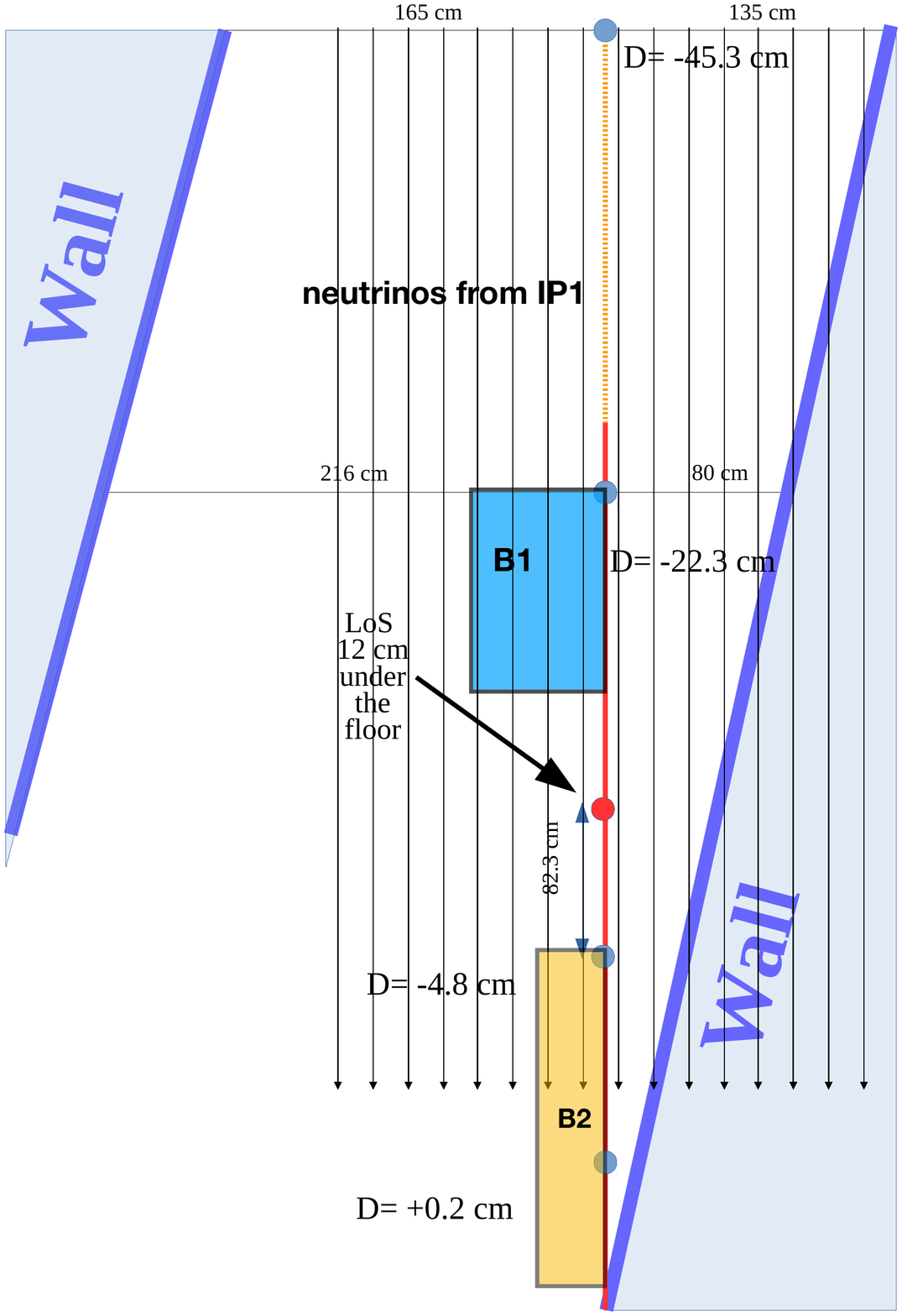}
    \includegraphics[width=0.45\textwidth] {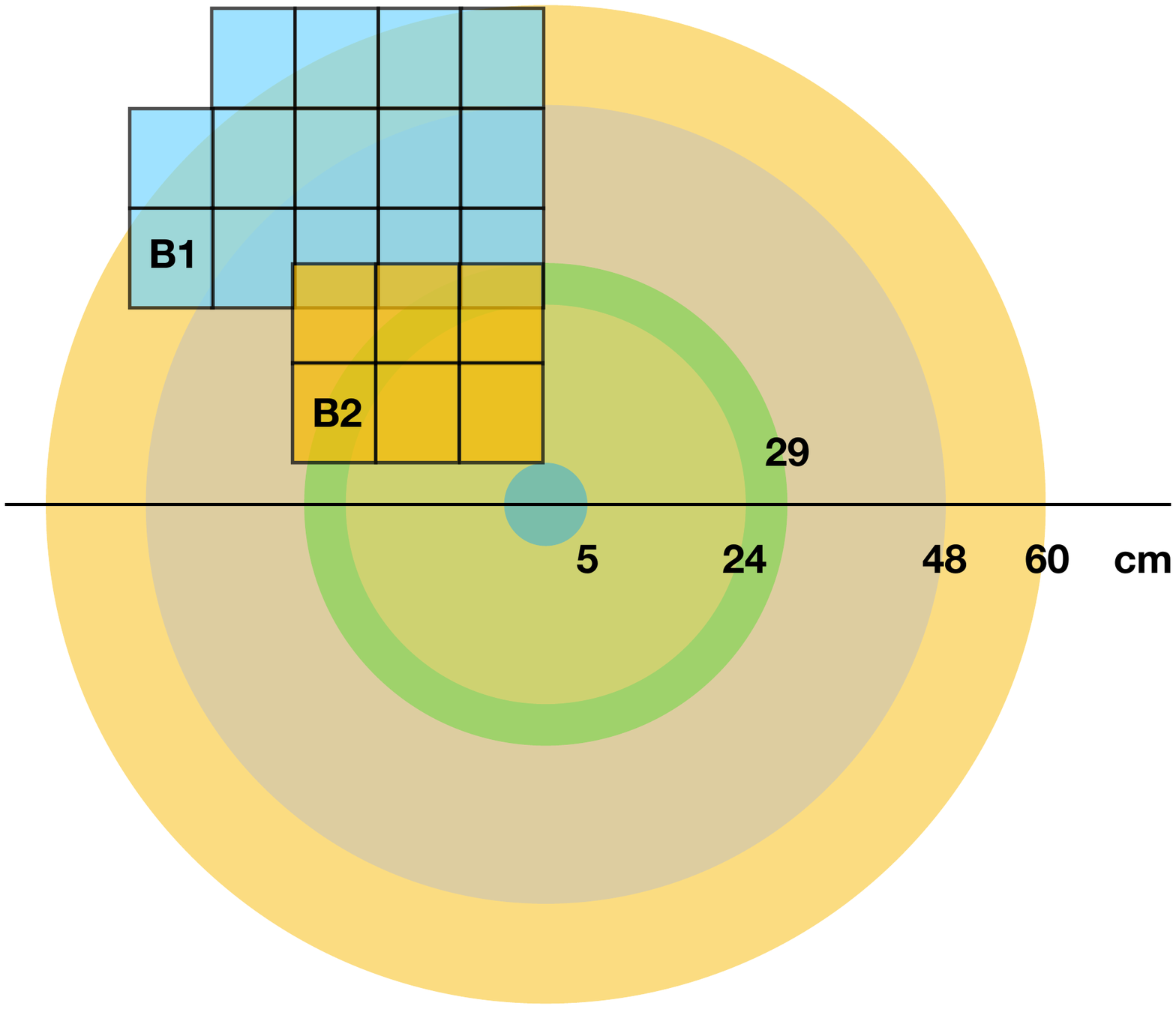}
\caption[B1B2 architecture] {A possible architecture of the B2 (Pilot ) and B1 (Phase2) detectors;
the numbering refers to the distance to the IP: B1 is closer. Top: 3D view of B2 
\cite{CERN-EN}. 
Bottom left: top view of B1 and B2; D is the height of the TI18 cavern floor from the LoS, measured in different points. Bottom right: section view of B1 and B2 showing their radial distance in cm from the beam centre, which is taken as 57.6 mm above the LoS.
 \label{fig:XSEN_B1B2}}
\end{figure}

In summary, 
there are two distinct situations: neutrinos emerging promptly from the IP ( c and b decays) and neutrinos from meson decays and secondary interactions. 
In the first situation the accuracy of the prediction depends on the physics simulation: a range of MonteCarlo programs and different PDFs can be compared, and an accuracy be estimated.  
In the second situation, the physics calculations are entangled with the LHC structure, its beam transport and cleaning elements, and with the running conditions (beam movements, unpredicted hot corners, for instance in our tests we observed sprays of hadrons from unexpected locations, etc…); it is difficult to determine the accuracy of the simulation, and real data can help. 
This is one of the motivations for the pilot run in 2021, and also for the architecture with
two independent detectors at different eta ranges, as shown in 
figure ~\ref{fig:XSEN_B1B2}.

\section{The Emulsion Detector for LHC Run3}

The preceding pages showed that there are parameters in the design of the XSEN experiment that can be optimized only through experience. Therefore we propose a plan in  two phases:\\

Phase1:\\
A compact detector of 0.4 to 1 ton to be installed in 2020 for a PILOT run to take data in 2021. A hundred high energy neutrino interactions can be collected. Experience will be acquired on: \\
-- local backgrounds. Their characterisation is input for tuning the simulations. Besides, from a measurement of the recorded track density we can establish for how long the emulsions can be exposed before reaching the performance limit, which we have estimated to be 30/fb in TI18; this impacts on the replacement rate, and on the cost of the experiment. \\
-- handling of emulsions.  Procurement times, transportation,  and the tools for brick assembling, disassembling and emulsion development. For easy logistics, the tools are to be located at CERN; those used for OPERA 
\cite{OPERA}
and for tests for SHIP 
\cite{SHIPtechnical} would be kindly available to us, but we need to understand their performance in a routine use and to study operation with personnel on shift.\\
-- analysis. Determine the brick fiducial volume: some  inefficiency is expected at the edges. The analysis sequence and tracking algorithms should be optimised for the environment. 
We have to understand the impact of the small variations in acceptance due to changes of the beam direction during a fill, and if we need to weight those with the ATLAS online luminosity information.  

Phase2:\\
The XSEN detector is extended to 1.5 and up to 3 tons, in two sections subtending two different eta ranges, for taking data from 2022 until the LHC Long Shutdown 3. Thanks to modularity and small size, the detector installation
can be performed in a week during the LHC winter shutdown 2021/22. 

\subsection{Detector Architecture and Predicted Event Statistics}

The basic unit of the setup is a stack made of thin layers of lead interleaved with nuclear emulsions.  The stack is then put into a custom packaging frame made of a plastic protective sheet, a plastic cover (both made of black polypropylene) and an aluminum structure, designed and constructed to be crimped onto the stack.  The object is then wrapped with an adhesive, low-outgassing aluminium tape (0.13 mm thick) which provides light tightness.  The last step also increases the mechanical stability and precision of the external surfaces.  The resulting "brick" is 128 mm wide, 104 mm high, 78 mm thick and weighs ~8.3 kg.

Figure ~\ref{fig:XSEN_B1B2}
shows a 3D view of a compact detector ( B2 ) for the  XSEN PILOT run. It consists of 108 bricks.  
Depending on its final structure, a brick can contain from 0.32 to 0.8 m$^2$ of emulsions.
Detector size and shape must be  optimized for performance and cost. 
At constant mass, the detector can be made wider and thinner, or narrower and thicker,
thus changing the neutrino $\eta$ acceptance and modifying the suppression efficiency for lower energy neutrino-N interactions.
The minimal implementation we are considering consists of 48 bricks (2x2x12) subtending the eta range 8.0$<\eta<$9.5.

Since the proton beams in IP1 cross vertically with a half-angle of 120 microradians, the beam is centred high above the LoS by 57.6 mm, and
the XSEN detector lies 108 mm above the LoS, slightly off beam axis. 
We expect that during Run3 the crossing half-angles in IP1 will change in the same range as during 2018.
That means that 
during fills the angles will go from 160 down to 120 microradians, 
corresponding to 76.8 mm and 57.6 mm of the LoS displacement at TI18 respectively. 

In Phase2 a second detector ( B1 ) is placed further uphill in TI18, towards IP1. 
Figure ~\ref{fig:XSEN_B1B2} shows an implementation with 168 bricks. Space allows for  doubling the azimuthal acceptance, if resources are available. 
The minimal configuration we are considering contains 96 bricks and has acceptance for 7.4$<\eta<$8.0.

\begin{figure} [h]
\centering
\includegraphics[width=0.6\textwidth]{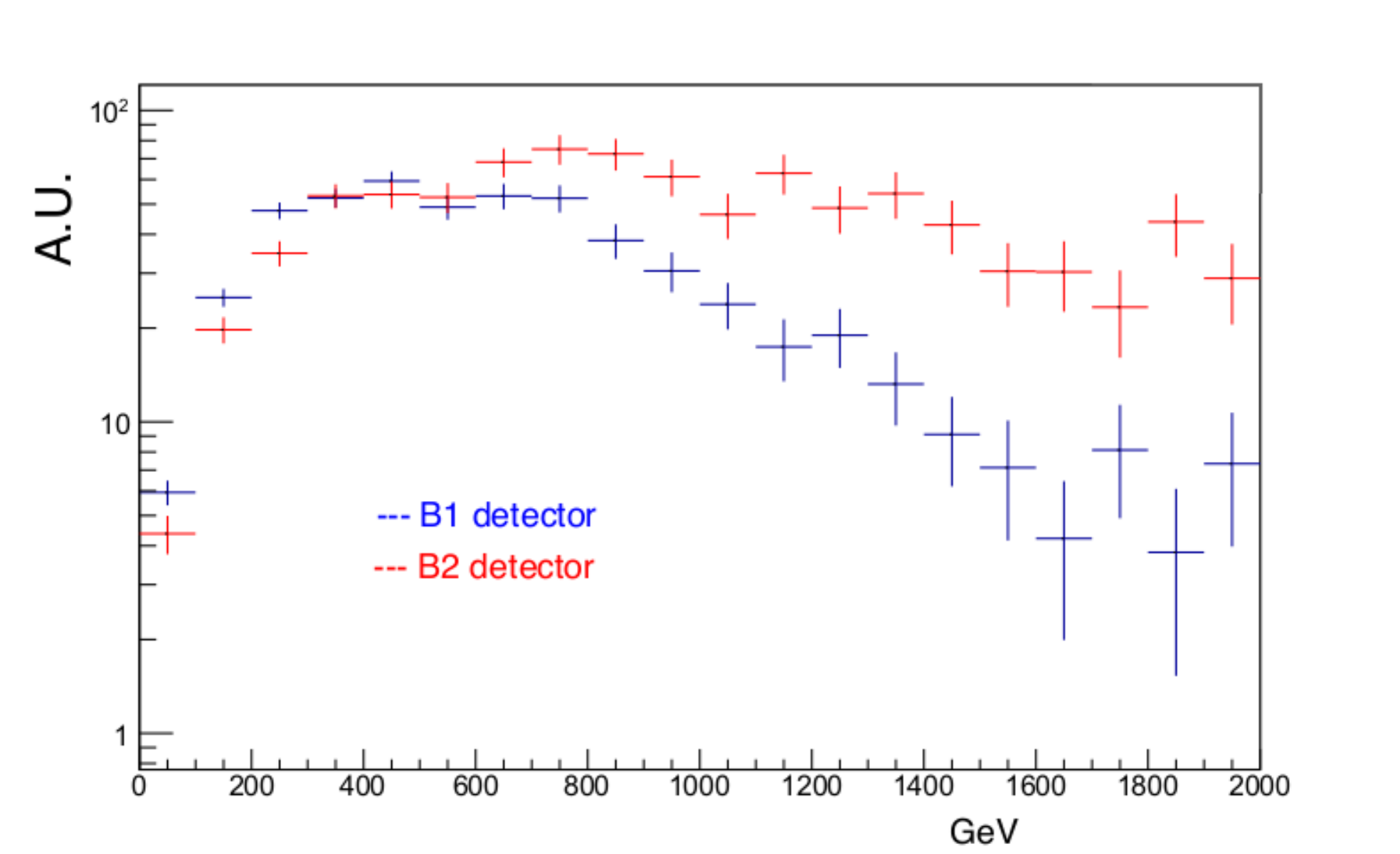}
\caption[spectrum] {Predicted energy spectra of neutrinos interacting in the B1 and B2 detectors.
\label{fig:DNDE_B2_B1}}
\end{figure}

Simulations of proton-proton interactions at 14 TeV were performed with Pythia.
An event by event weight was applied for properly taking into account the $\nu$N interaction cross-section dependence on energy and on neutrino flavour
\cite{nutauN, PDG}. 
Table  
~\ref{tab:eventrate}  
summarizes the expected statistics and characteristics of the events for the detector configuration shown in Fig.
~\ref{fig:XSEN_B1B2}.
For estimating the uncertainties on the cross-section measurement, we considered an efficiency of 50\% and a systematic error as large as the statistical one.
The predicted energy spectrum of neutrino events in both the B1 and B2 detectors are shown in Figure
~\ref{fig:DNDE_B2_B1}.
The additional contribution of pion and Kaon decays is not included (see discussion on neutrino flux).
\begin{table} [h]
\begin{center}
\topcaption{ Expectations for a detector configuration as shown in Figure  
~\ref{fig:XSEN_B1B2}. B2 and B1 are made of 108 and 168 bricks respectively.}
\label{tab:eventrate}
\begin{tabular} {lrrrrr} \hline 
& & & & \\
  &B2 Pilot  &B2  &B1 &B2+B1  &B2+2xB1  \\
   &25 /fb &150  /fb &150 /fb  &150 /fb  &150 /fb\\ \hline
& & & & \\
integral $\nu$ fluence  &5.6$\times$10$^{10}$ &3.4$\times$10$^{11}$ &4.6$\times$10$^{11}$   &0.8$\times$10$^{12}$  &1.3$\times$10$^{12}$    \\
all flavour $\nu$ events &142 &852 &490   &1342  &1832    \\
tau flavour $\nu$  events   &4  &25 &26   &51  &77  \\ 
$\eta$ range &8.0-9.5 &8.0-9.5 &7.4-8.2 & & \\
average E$_{\nu}$ (RMS) GeV &1200(600) &1200(600) &700(400) & &  \\
$\Delta\sigma_{\nu N}$ &17\% &7\% &9\% &6\% &5\%  \\
& & & & \\ \hline
\end{tabular}
\end{center}
\end{table}

\subsection{Emulsion Handling}
\label{sec:emu_handl}
\begin{figure}[htbp]
\centering
\includegraphics[scale=0.2]{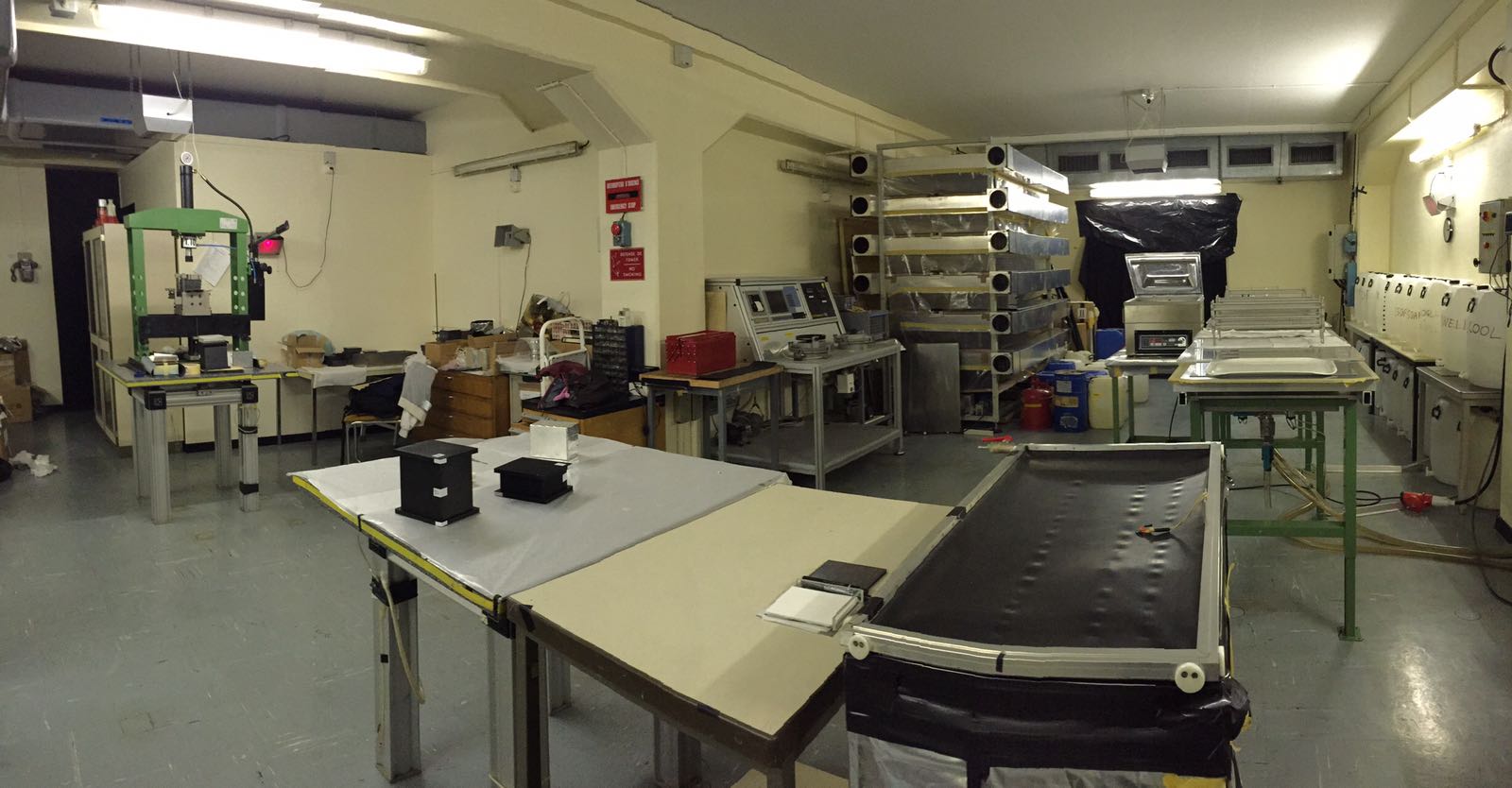}
\caption{The SHiP
\cite{SHIPtechnical}
 emulsion laboratory at CERN.}\label{fig:emulsionfacility}
\end{figure}

The preparation of the emulsion targets and the subsequent development of emulsion films require a dark room equipped with dedicated facilities. The SHiP 
\cite{SHIPtechnical}
emulsion laboratory located at CERN Meyrin Site in building 169 (Fig.
~\ref{fig:emulsionfacility})
is suited for that purpose and dimensioned for handling the amount of emulsions required for the pilot run in 2021.  It hosts two different areas: one room is used for the assembling of the target units and another one is equipped with a  development line with the corresponding temperature control to keep the solutions at the required temperature. Dedicated humidifiers and conditioners 
guarantee a stable (20$\pm$1)$^\circ$C temperature and a (60$\pm$2)\% humidity level necessary to handle the emulsion films. 

\begin{figure}
\centering
\includegraphics[scale=0.4]{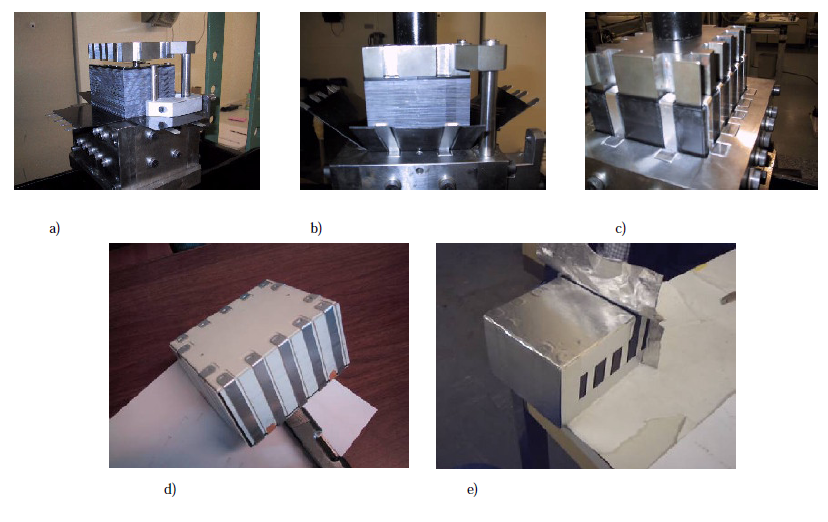}
\caption{Sequence of the spider packaging procedure.}\label{spider_procedure}
\end{figure}

The assembly of emulsion films and passive layers will be performed using the packaging procedure adopted in the OPERA experiment and commonly referred to as `spider packaging procedure'. It is based on a 800~$\mu$m thin aluminum foil, called `spider' that provides mechanical stability to emulsion films and passive layers, which are stacked together to form a pile. The spider is firstly placed under the pile (Fig. \ref{spider_procedure}a), then it is folded on the sides by mechanical pressure (Fig. \ref{spider_procedure}b) and closed on the upper emulsion film (Fig. \ref{spider_procedure}c). Plastic side protection and cover keep the rigidity and avoid the direct contact between emulsions and aluminum (Fig. \ref{spider_procedure}d), the light shielding is provided by wrapping an adhesive aluminum tape around the pile (Fig. \ref{spider_procedure}e).

The development process makes the latent image visible by the reduction of silver
ions to metallic silver. It consists of five steps:
\begin{itemize}
    \item {\it Development}: The solution is made with Fuji Developer and Fuji Starter in demineralised water. Chemical agents reduce the crystals containing a latent image center while leaving the remnants unchanged. The emulsions are left in this solution for 25 minutes.   
    \item {\it Stop}: To stop immediately the development process and control precisely the time, a bath with acetic acid dissolved in water is used. As a matter of fact, the action of chemical developers is strongly dependent on pH: the less alkaline the environment, the less active the developer is. This step lasts 10 minutes.
    \item {\it Fix}: The fixation bath is a solution of Fuji UR-F1 (Acetic Acid, Sodium Thiosulphate, Sodium Acetate, Sodium Sulfite and Ammonium Thiosulfate) in water. It removes all the
residual silver halide, leaving the metallic silver to form the image. After one hour in this bath a check of the transparency of the emulsion films is performed. If the check gives a
positive result, it is possible to move to the last step of the development line, otherwise, the emulsions are left in the fixation bath and checked every 20 minutes until the desired
transparency is reached.
      \item {\it Wash}: To remove all the silver thiosulphate complexes in the emulsion, which could otherwise obscure the image, the films are washed in circulating water for 80 minutes.
  \end{itemize}
  
      At the end of the development process, the emulsions are left to dry for 24 hours before performing a glycerine treatment. Indeed, after the development process, the emulsion layers
reach a thickness of 30$\mu$m because of the empty spaces left from the removal of the silver halides. This last step is necessary to inflate the emulsion back to their original thickness of 70$\mu$m 
to avoid the shrinkage effect on track slopes. The emulsion films are left for two hours in a bath of 20\% of glycerine and water. In this way the water penetrates the emulsion layer and the glycerine
fills the holes left from the silver halides. In the end, the emulsions are dried at a constant humidity level of (60$\pm$2)\% for two days.

\subsection{Emulsion Scanning and Analysis}

\begin{figure}[htbp]
\centering
\includegraphics[scale=0.3]{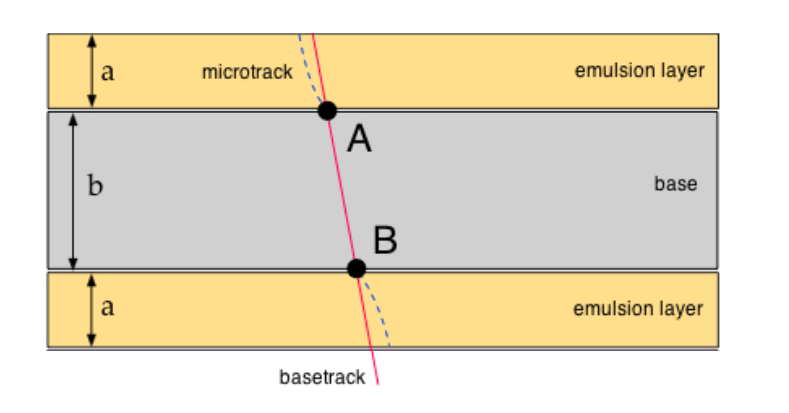}
\caption{Layout of an emulsion film. Two 70 $\mu$m-thick layers of nuclear emulsions are separated by a 175~$\mu$m-thick plastic base.}\label{ship_film}
\end{figure}

The proposed emulsion films for this experiment consist of two 70$\mu$m-thick layers of nuclear emulsion, separated by a 175~$\mu$m-thick plastic base (Fig. \ref{ship_film}). The transverse size is $12.8 \times 10.4$~cm$^2$, like for the passive plates.\\ 

The track left by a charged particle in an emulsion layer is recorded by a series of sensitized AgBr crystals, growing up to 0.6 $\mu$m diameter during the development process. An automated optical microscope analyses 
the whole thickness of the emulsion, acquiring various topographic images at equally spaced depths. The acquired images are digitized, then an image processor recognizes the grains as  \textit{clusters}, i.e.~groups of pixels of given size and shape. Thus, the track in the emulsion layer (usually referred to as \textit{microtrack}) is obtained connecting clusters belonging to different levels. Since an emulsion film is formed by two emulsion layers, the connection of the two microtracks through the plastic base provides a reconstruction of the particle's trajectory in the emulsion film, called \textit{base-track}.

The analysis of emulsion films is performed using new generation optical microscopes. For the OPERA experiment two different scanning systems were developed: one in Japan, the Super Ultra Track Selector (S-UTS) \cite{Morishima:2010zz}, and one by a collaboration of the different European laboratories, the European Scanning System (ESS) \cite{Armenise:2005yh, Arrabito:2006rv, Arrabito:2007td, DeSerio:2005yd}. The ESS is a microscope equipped with a computer-controlled monitored stage, movable along both X and Y axes and in the Z direction, a dedicate optical system and a CMOS Mega-pixel camera mounted on top of the optical tube. For each field of view, it executes the following steps:

\begin{itemize}
\item local tomography;
\item cluster recognition;
\item grain selection;
\item three-dimensional reconstruction of aligned cluster grains;
\item parameter extraction for each grain sequence.
\end{itemize}
The ESS allows the scanning of the whole emulsion  surface with a maximum speed of 20 cm$^2$/h.

An upgrade of the ESS system was  performed by the Naples emulsion scanning group \cite{Alexandrov:2016tyi}. The use of a faster camera with smaller sensor pixels and a higher number of pixels combined with a lower magnification objective lens, together with a new software LASSO \cite{Alexandrov:2016tyi,Alexandrov:2015kzs} has allowed to increase the scanning speed 
to 190 cm$^2$/h, 
one order of magnitude larger than before. 
The lens of the microscope guarantees a submicron resolution and, having a working distance
in Z longer than 300 $\mu$m, to scan both sides of the emulsion film. To make the optical path homogeneous in the film, an immersion lens in an oil with the same refraction index of the emulsion is used. A single field of view is 800$\times$600 $\mu$m$^2$; larger areas are scanned by repeating the data acquisition on a grid of adjacent fields of view. The images grabbed by the digital camera are sent to a vision processing board in the control workstation to suppress noise.
Three-dimensional sequences of aligned clusters (grains) are recognised and reconstructed by the CPUs of the host workstation. The track recognition procedure is a quite complex process executed by the LASSO software tracking module. The steps of the algorithm for the {\it micro-track} reconstruction are described in \cite{Alexandrov:2015kzs}.
Micro-tracks are identified by a sequence of aligned grains. The position assigned to a micro-track is its intercept with the nearest plastic base surface. Once micro-tracks have been reconstructed, the following steps of the analysis are performed
with a dedicated offline software \cite{Tyukov:2006ny}. After development, the thickness of the emulsion layer turns out to be reduced (shrinkage) due to the dissolution of silver halides in the fixing phase. The glycerin treatment described in sec.~\ref{sec:emu_handl} corrects this effect, but only on large scales. Therefore, as a first step, a procedure aiming at a finer correction of the shrinkage is applied. Subsequently, the two corresponding track segments in either emulsion layers are linked forming a so-called  {\it base-track}: this step is called linking. This is important to reduce the instrumental background due to fake combinatorial alignments and to increase the precision on the reconstruction of the
track angle, minimising distortion effects. The base-tracks are formed linking the closest grains to the plastic base in the two different emulsion layers and are selected on the basis of a $\chi^2$ cut. The full-volume wide reconstruction of particle tracks requires connecting base-tracks in consecutive films. In order to define a global reference system, a set of affine transformations has to be computed to account for the different reference frames used for data taken in different
plates.

\subsection{Installation and Refurbishing}

The bricks are rather robust objects, easy to transport and handle.
The number of required bricks is relatively small;
more bricks can be put together in a single assembly by using a manual press in the existing dark room facility in CERN building 169.
Sets of ~40-50 bricks are small and light enough to be easily lowered into the LHC tunnel in a lift for personnel and  moved around in small wooden carts.  Depending on the chosen detector design several such carts will be required for transport of the complete set of emulsion bricks.
The installed detector will be protected with a layer of neutron attenuator, either 3-9 cm of borated polyethylene or a foam of boron carbide (\`a la NA62) 
\cite{NA62}.

Tolerances of the brick sizes ($\pm$ 0.5 mm) make their placement with precision higher than 1 mm difficult.  However, special attention will be given to recording the placement of individual bricks in the setup, so that distance from the line of sight is well known.  Survey targets are foreseen on the frame of the support platform so that the brick positions can be correlated with the accelerator coordinate system.

\begin{figure}[htbp]
\centering
\includegraphics[scale=0.55]{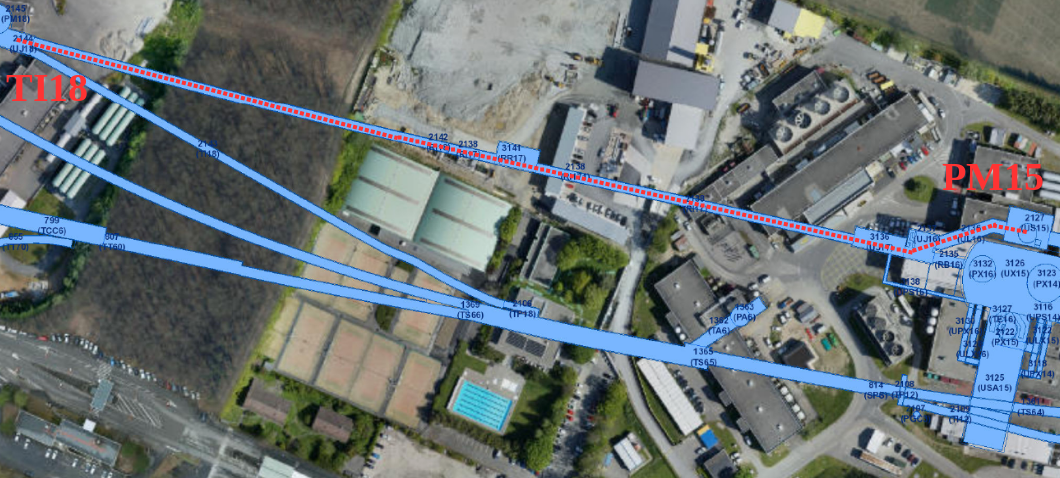}
\caption{Access path to the proposed location for the XSEN detector in the TI18 cavern.}\label{PM15_TI18}
\end{figure}

Access from the surface into the LHC tunnel for installation and removal will be performed through the PM15 shaft
(Fig. \ref{PM15_TI18}).  
Another nearby access point, PM18, does not allow transport of a few hundred kilograms of material.
Transport path along the LHC tunnel and access to TI18 cavern are on the opposite sides of the cryogenic lines. Transfer of any material will have to be done in a way that does not jeopardize the integrity of either the cryogenic line or the accelerator itself.  Therefore we planned to profit from the small size of the boxes and transfer them under the cryogenic line at the level of MBA13.R1.  Such a transfer has to be limited to only a predefined area free of any obstacles on the floor. 
If required an appropriate mechanical protection may be installed to protect the cryogenic line from unintended contacts.
An access test was performed by D. Lazic together with  M. Lamont and with P. Santos Diaz (CERN EN). They went along the whole path between PM15 and TI18 to check if there is a problem.  The lift is the big one with capacity of 3000 kg and the whole path from the lift PM15, through US15, through bypass UL15 and the LHC tunnel, to TI18, is 1.5 m wide, flat and without any obstacles. 
For the passage under the cryogenic line near the TI18, 
it was found that even in the most conservative choice of the place where the material is passed under the beam line we can have a passage that is ~80 cm wide and 34 cm high.

Transfer of people to the other side of the beamline can be easily done thanks to an already installed bridge over MBB13.R1.

The project was presented at the CERN TREX committee. No showstopper was identified and no safety issues found.\\ 
After the extraction of the emulsions from the tunnel they should be developed as quickly as possible.  Therefore, good communication lines with Radioprotection and Transport teams have to be put in place so that the emulsions are transported to the dark room on Meyrin side as soon as all the required procedures are performed.
Photographic emulsions are slightly flammable but they are interleaved with lead sheets and enclosed in an aluminum foil. The same is true for the neutron shielding.  The detector is completely passive, there is no power source anywhere near it. In the past measurements it was sufficient to obtain a permission from HSE to install the emulsions.
There are no gases or liquids of any kind in the setup. Liquids used in the development lab are dealt through the usual CERN channels for chemical waste.

Activites in the TI18 can be divided in two distinct categories:

a) Installation of the support structure(s) and alignment with respect to the accelerator coordinate system should take place once the setup has reached a reasonable level of maturity.  The required equipment is rather simple - an electric drill that can drill into the concrete, a vacuum cleaner from Radioprotection and few hand tools for bolting the support structure into the floor.  The support structure itself will be made in a way that allows easy passage under the beamline and a quick assembly in its final place.  The preferred time for this operation is before the beamline is filled with helium, so before august 2020 in the current planning.

b) We expect that the installation of the individual bricks will take less than a minute per brick.  Even when we add the time needed to move the bricks in and out of TI18, set the portable light and ventillation, record individual stages of the installation and finally remove all the ancillary equipment it amounts to less than eight hours of work for two persons.  Once LHC reaches its full luminosity in 2022 we plan to exchange emulsions during Technical Stops, i.e. after ~30/fb of integrated luminosity.  Speed and simplicity of the exchange process gives us confidence that the operation can be done without interfering with the LHC schedule.

Important aspect is that any work on the setup takes place inside TI18, so once the equipment is safely transported, there is no conflict with any other activity in the tunnel.

\bibliography{auto_generated}   

\end{document}